\documentclass[english,aps,twocolumn,amsmath,amssymb,superscriptaddress,pre]{revtex4-2}
\usepackage[T1]{fontenc}
\usepackage[latin9]{inputenc}
\usepackage{babel}
\usepackage{array}
\usepackage{verbatim}
\usepackage{float}
\usepackage{mathtools}
\usepackage{dsfont}
\usepackage{amsmath}
\usepackage{amssymb}
\usepackage{graphicx}
\usepackage[unicode=true,pdfusetitle,
bookmarks=false,
breaklinks=false,pdfborder={0 0 1},backref=false,colorlinks=false]
{hyperref}
\usepackage{algorithm}
\usepackage{algpseudocode}

\makeatletter

\providecommand{\tabularnewline}{\\}
\floatstyle{ruled}
\newfloat{algorithm}{H}{loa}
\providecommand{\algorithmname}{Algorithm}
\floatname{algorithm}{\protect\algorithmname}

\newcommand{\ox}{\overline x}

\DeclareMathOperator{\sign}{sign}

\makeatother

\begin{document}
	
	\title{Matrix Product Belief Propagation for reweighted stochastic dynamics over graphs}
	\author{Stefano Crotti}
	\email{stefano.crotti@polito.it}
	
	\affiliation{Department of Applied Science and Technology, Politecnico di Torino, 10129, Turin, Italy}
	\author{Alfredo Braunstein}
	\affiliation{Department of Applied Science and Technology, Politecnico di Torino, 10129, Turin, Italy}
	\affiliation{Italian Institute for Genomic Medicine, 10126, Turin, Italy}

\begin{abstract}
Stochastic processes on graphs can describe a great variety of phenomena ranging from neural activity to epidemic spreading. While many existing methods can accurately describe typical realizations of such processes, computing properties of extremely rare events is a hard task. Particularly so in the case of recurrent models, in which variables may return to a previously visited state. Here, we build on the matrix product cavity method, extending it fundamentally in two directions: first, we show how it can be applied to Markov processes biased by arbitrary reweighting factors that concentrate most of the probability mass on rare events. Second, we introduce an efficient scheme to reduce the computational cost of a single node update from exponential to polynomial in the node degree. Two applications are considered: inference of infection probabilities from sparse observations within the SIRS epidemic model, and the computation of both typical observables and large deviations of several kinetic Ising models.
\end{abstract}
\maketitle

The problem of computing observables and marginal probabilities on
a complex Markov process on large networks has been addressed extensively
in the literature. While Monte-Carlo procedures can be often effective
to compute averages approximately, they suffer from two separate issues:
large relative sampling errors when computing averages that cancel
out at the first order and they are limited to sampling ``typical''
events, as nontypical ones require an exponential number of samples.
To address the first issue, many analytical solutions, mainly based
on mean-field methods, have been devised \citep{van2008virus,karrer2010message,delferraro2015dynamic,pelizzola2013variational,pelizzola2017variational,aurell2017cavity,ortega2022dynamics}.
A solution that is exact on acyclic graphs is Dynamic Cavity (DC)
\citep{neri2009cavity}. DC on general processes suffers from one
main drawback, the fact that one must be able to represent the joint
distribution of a single variable trajectory and a feedback field,
and with some exceptions, the space of these trajectories is exponentially
large (in the time horizon), and thus the approach becomes impracticable.
One of these exceptions is on ``non-recurrent'' models, i.e. models
in which each variable can only progress sequentially through a finite
set of $k$ states, never going back to a previous state. In these
cases the set of trajectories is polynomial in the time horizon (as
an example with $q=3$, a trajectory $\left(1,1,2,2,2,2,3,3\right)$
on epochs $t=0,\dots,7$ can be represented by the integer tuple $\left(2,6\right)$
of epochs on which the variable effectively progresses to the next
state in the sequence). Examples of non-recurrent models are the SI,
SIR, SEIR compartmental models in computational epidemics, in which
an individual can only transition from Susceptible to Exposed, from
Exposed to Infective and from Infective to Recovered. While the use
of non-recurrent models is pervasive, oftentimes a more realistic
description demands that re-infections be taken into account. In such
cases, ``recurrent'' models such as the SIS and SIRS are employed.
Additionally, important processes in statistical physics such as Glauber
dynamics belong to the class of models with recurrence.

In a recent work \citep{barthel2018matrix,barthel2020matrix}, an
interesting DC variant was proposed that exploits the Matrix Product
State representation (MPS) to parametrize site trajectories and applied
it to the Glauber dynamics on a Random Regular (RR) graph with degree
3. While these results are promising, the scheme suffers from two
major limitations: first, it is computationally expensive (the update
on a node of degree $z$ is of the order of $M^{2z-1}$ \citep{barthel2020matrix}
where $M$ is the matrix dimension), making it impractical even for
moderately large Erdos-Renyi (ER) random graphs, in which some large-degree
vertices are surely present. Second, the scheme is devised to analyze
a ``free'' dynamics without any sort of reweighting, which as we
will see is necessary to study atypical trajectories. Matrix Product
States, also known as Tensor Trains, are not new in physics and other
areas of science, as they have been successfully applied both in many-body
quantum systems \citep{perez2007matrix,verstraete2006matrix,fannes1992finitely},
out-of-equilibrium statistical physics \citep{derrida1993exact,banuls2019using},
machine learning \citep{han2018unsupervised,stoudenmire2016supervised}
and more.

We propose an alternative approach, dubbed Matrix Product Belief Propagation
(MPBP), based on the Pair Trajectory Belief Propagation formulation
which was first introduced in \citep{altarelli_large_2013}. It is
closely related to DC but allows naturally to include non-negative
reweighting terms on stochastic trajectories, thus allowing to study
large deviations of the system. In practical terms, MPBP consist on
a fixed point equation that is solved by iteration, whereas DC is
solved sequentially in time, with a number of steps which is equal
to the number of epochs of the dynamics. The latter approach is inherently
limited to free dynamics: building trajectories sequentially in time
makes it impossible to account for the effect of reweighting terms
relative to future epochs.

The Julia code used to implement the method and produce the data presented
in this work is publicly accessible at \citep{repo}.

We describe in the following the models under consideration. Given
a graph $G=\left(V,E\right)$ with $V=\left\{ 1,\dots,N\right\} $,
consider a joint distribution over a set of discrete variables $\boldsymbol{x}=\{x_{1},\ldots,x_{N}\}$
throughout $T$ successive epochs of the form
\begin{equation}
p(\overline{\boldsymbol{x}})=\frac{1}{Z}\prod_{t=0}^{T-1}\prod_{i=1}^{N}f_{i}^{t+1}\left(x_{i}^{t+1},\boldsymbol{x}_{\partial i}^{t},x_{i}^{t}\right).\label{eq:p_general}
\end{equation}

We use bold letters to indicate multiple variable indices $\boldsymbol{x}_{A}\equiv\{x_{j}\}_{j\in A}$
and overbars for multiple times indices $\overline{x}\equiv\{x^{t}\}_{t=0:T}$.
Moreover, we indicate by $\partial i=\left\{ j:\left(ij\right)\in E\right\} $
the set of neighbors of index $i$.

The form (\ref{eq:p_general}) includes (but notably is more general
than) reweighted Markov dynamics $f_{i}^{t+1}(x_{i}^{t+1},\boldsymbol{x}_{\partial i}^{t},x_{i}^{t})=w(x_{i}^{0})^{\delta(t,0)}w(x_{i}^{t+1}|\boldsymbol{x}_{\partial i}^{t},x_{i}^{t})\phi_{i}^{t+1}(x_{i}^{t+1})$
with stochastic transitions $w$ and reweighting factors $\phi$
\begin{equation}
p(\overline{\boldsymbol{x}})=\frac{1}{Z}\prod_{i=1}^{N}w(x_{i}^{0})\prod_{t=0}^{T-1}w(x_{i}^{t+1}|\boldsymbol{x}_{\partial i}^{t},x_{i}^{t})\phi_{i}^{t+1}(x_{i}^{t+1}).\label{eq:p}
\end{equation}
$\delta(y,z)$ is the Kroenecker delta which evaluates to $1$ if $y=z$, to $0$ otherwise, and $w(x_{i}^{0})$ is the initial state probability, which we take to be factorized over the sites. 

Note that $Z=1$ in the absence of reweighting factors. Two
types of reweighted dynamics of the form (\ref{eq:p}) will be used
as running examples throughout this work. The first is Bayesian inference
on a process of epidemic spreading. The posterior probability of the
epidemic trajectory $\overline{\boldsymbol{x}}$ given some independent
observations $\{O_{i}^{t}\}$ on the system is given by 
\begin{align}
p(\overline{\boldsymbol{x}}|O) & =\frac{1}{p(O)}p(\overline{\boldsymbol{x}})p(O|\overline{\boldsymbol{x}}).\label{eq:bayes}
\end{align}
\eqref{eq:bayes} can be seen as a particular case of \eqref{eq:p}, where $p(\overline{\boldsymbol{x}})=\prod_{i=1}^{N}w(x_{i}^{0})\prod_{t=0}^{T-1}w(x_{i}^{t+1}|\boldsymbol{x}_{\partial i}^{t},x_{i}^{t})$ and corresponds to the distribution of the free dynamics of the chosen epidemiological
model, $p(O|\overline{\boldsymbol{x}}) = \prod_{i}\prod_t p(O_{i}^{t}|x_{i}^{t})=\prod_{i}\prod_t\phi_{i}^{t}(x_{i}^{t})$ and $Z=p(O)$.
 % &\frac{1}{Z}\prod_{i=1}^{N}w(x_{i}^{0})\prod_{t=0}^{T-1}w(x_{i}^{t+1}|\boldsymbol{x}_{\partial i}^{t},x_{i}^{t})p(O_{i}^{t}|x_{i}^{t})

The simplest among the recurrent epidemiological models is
the Susceptible-Infectious-Susceptible (SIS), where each individual
starts with a probability $\gamma_{i}$ of being infectious at time
zero. Then at each time step a susceptible node $i$ can be infected
by each of its infectious neighbors $j\in\partial i$ with probability
$\lambda_{ji}$, and an infectious node can recover with probability
$\rho_{i}$. Observation terms $p(O_{i}^{t}|x_{i}^{t})$ are naturally
used to model medical tests: $O_{i}^{t}$ is the outcome of a test
performed on individual $i$ at time $t$. This formalism allows to
incorporate information about the degree of accuracy of tests.

The second example is parallel Glauber dynamics for an Ising model
at inverse temperature $\beta$ with couplings $\{J_{ij}\}$ and external
fields $\{h_{i}\}$. Besides being one of the paradigmatic models
in theoretical non-equilibrium statistical physics, Glauber dynamics
is employed in the study of neural activity \citep{renart2010asynchronous,roudi2011mean}.
It is defined by transitions
\begin{equation}
\tilde{w}(\sigma_{i}^{t+1}|\boldsymbol{\sigma}_{\partial i}^{t})=\frac{e^{\beta\sigma_{i}^{t+1}\left(\sum_{j\in\partial i}J_{ij}\sigma_{j}^{t}+h_{i}\right)}}{2\cosh\left[\beta\left(\sum_{j\in\partial i}J_{ij}\sigma_{j}^{t}+h_{i}\right)\right]}.\label{eq:w_glauber}
\end{equation}

The dynamics does not converge to the equilibrium of the underlying Ising model $p_{J,h}(\boldsymbol{\sigma})=Z^{-1}\exp[-H_{J,h}(\boldsymbol{\sigma})]$, but it allows to compute observables of interest in some cases (see the Supplementary Information).

Moreover, we will allow $\sigma_i$ to stay in the same state with probability
$p_{0}$. The transition thus becomes
\begin{align}
w(\sigma_{i}^{t+1}|\boldsymbol{\sigma}_{\partial i}^{t},\sigma_{i}^{t})=&(1-p_{0})\tilde{w}(\sigma_{i}^{t+1}|\boldsymbol{\sigma}_{\partial i}^{t})+\nonumber\\
&+p_{0}\delta(\sigma_{i}^{t+1},\sigma_{i}^{t}).\label{eq:glauber2}
\end{align}

In the limit $p_0\to 0$, the stationary distribution converges to $p_{J,h}$ because the dynamics reduces to an asynchronous one (two or more simultaneous state changes happen with probability $\mathcal{O}(p_0^2)$). See also the Supplementary Information.

Additionally, such dynamics can be ``tilted'' with e.g. a term $\prod_{i}\phi_{i}^{T}(\sigma_{i}^{T})=\prod_{i}e^{h\sigma_{i}^{T}}$
in order to study atypical trajectories. Note that other models studied
in physics such as Bootstrap Percolation can be remapped into Glauber
dynamics \citep{ohta2010universal}. 

\subsection*{Related work}

\paragraph{Mean-field methods}

We briefly review the main features of existing approaches based on
the cavity method. Dynamic Message Passing (DMP) \citep{karrer2010message,van2011n,shrestha2015message}
and the Cavity Master Equation \citep{ortega2022dynamics,aurell2017cavity}
are simple and fast approximate methods that were originally formulated
on continuous time as ODEs for a vector of single-edge quantities
(such as cavity magnetizations). Both methods are exact on acyclic
graphs on non-recurrent models (such as SI or SIR), but only approximate
on non-non-recurrent ones, and do not allow for atypical trajectories.
$n$-step Dynamic Message Passing \citep{delferraro2015dynamic} makes
an $n$-Markov \emph{ansatz} on messages, exploring mainly $n=1$;
its features are essentially those of DMP, with the difference that
it applies to discrete time evolution and describes explicitly interactions
at distance $n$ in time. Different flavors of the cluster variational
method \citep{pelizzola2017variational,vazquez2017simple} approximate
the dynamics by treating exactly correlations between variables that
are close either in time or space. Large deviations have been studied
in \citep{del2014perturbative} using a perturbation theory in the
particular case of Glauber dynamics on a chain. Table \ref{tab:comparison}
summarizes the features of the methods mentioned above. We take into
consideration: ability to deal with reweighted dynamics, to
deal with recurrent models, and to compute autocorrelations at arbitrary
(time) distance.
\begin{table*}
\renewcommand{\arraystretch}{1.5}
\begin{centering}
% \begin{tabular*}{0.8\textwidth}{@{\extracolsep{\fill}}|>{\centering}m{0.3\textwidth}|>{\centering}m{0.12\textwidth}|>{\centering}m{0.16\textwidth}|>{\centering}m{0.16\textwidth}|}
\begin{tabular*}{0.9\textwidth}{@{\extracolsep{\fill}}|>{\centering}m{0.35\textwidth}|>{\centering}m{0.12\textwidth}|>{\centering}m{0.16\textwidth}|>{\centering}m{0.16\textwidth}|}
\hline 
 & reweighting & Recurrent models & Autocorrelations\tabularnewline
\hline 
BP for non-recurrent models \citep{altarelli_large_2013} & Y & N & Y\tabularnewline
\hline 
IBMF \citep{van2008virus}, DMP \citep{karrer2010message,shrestha2015message,delferraro2015dynamic},
CME \citep{aurell2017cavity} & N & Y & N\tabularnewline
\hline 
Dynamic Cluster Variational \citep{pelizzola2017variational} & {*} & Y & Only two-times\tabularnewline
\hline 
Matrix Product Dynamic Cavity \citep{barthel2018matrix} & N & Y & Y\tabularnewline
\hline 
Matrix Product Belief Propagation & Y & Y & Y\tabularnewline
\hline 
\end{tabular*}
\par\end{centering}
\medskip

\centering{}\caption{\label{tab:comparison}Features of existing analytical methods for
the description of stochastic dynamics on graphs, Y for yes, N for
no. The asterisks mean that the method could in principle be extended
to include the considered feature although this has not, to the best
of our knowledge, been done in the literature. IBMF stands for Individual-Based
Mean Field, DMP for Dynamic Message Passing, CME for Cavity Master
Equation. We did not include the perturbative approach \citep{del2014perturbative}
because it focuses on a very particular setting. }
\end{table*}

\paragraph{Monte Carlo}

Throughout this work, the performance of algorithms is compared with
Monte Carlo simulations. To estimate observables in a reweighted dynamics
of the form (\ref{eq:p}) we employ a weighted sampling technique
(see e.g. \citep{antulov2015identification}): the posterior average
of an observable $f$ is approximated by
\begin{equation}
\hat{f}=\frac{\sum_{\mu=1}^{M}\prod_{i,t}\phi_{i}^{t}\left((x_{i}^{t})^{(\mu)}\right)f\left(\boldsymbol{\overline{x}}^{(\mu)}\right)}{\sum_{\mu=1}^{M}\prod_{i,t}\phi_{i}^{t}\left((x_{i}^{t})^{(\mu)}\right)}
\end{equation}
where $\left\{ \boldsymbol{\overline{x}}^{(\mu)}\right\} _{\mu}$
are $M$ independent samples drawn from the prior $\prod_{i=1}^{N}w(x_{i}^{0})\prod_{t=0}^{T-1}w(x_{i}^{t+1}|\boldsymbol{x}_{\partial i}^{t},x_{i}^{t})$.
Such strategy, however, turns out to be computationally prohibitive
whenever the reweighting terms $\phi$ put most of the probability
mass on atypical trajectories, which are (exponentially) unlikely
to ever be sampled.

\subsection*{Matrix Product Belief Propagation}

For the dynamic version of Belief Propagation (BP), we start with \eqref{eq:p_general} as a distribution for single site trajectories $\overline{x}_{i}$. The associated factor graph would present many small loops due to the presence of both $\overline{x}_{i}$ and $\overline{x}_{j}$ in factors $f_i$ and $f_j$. Therefore, we
work directly on the so-called dual factor graph where variables are pair of
trajectories $(\overline{x}_{i},\overline{x}_{j})$ living on the
edges of the original graph. For more details about this step we refer
the reader to \citep[fig. 3, eqns 8,9]{altarelli_large_2013}. The
BP equations on the dual factor graph read
\begin{align}
m_{i\to j}(\overline{x}_{i},\overline{x}_{j})\propto & \sum_{\overline{x}_{\partial i\setminus j}}\prod_{t=0}^{T-1}f_{i}^{t+1}(x_{i}^{t+1},\boldsymbol{x}_{\partial i}^{t},x_{i}^{t})\nonumber \\
 & \times\prod_{k\in\partial i\setminus j}m_{k\to i}(\overline{x}_{k},\overline{x}_{i}).\label{eq:bp}
\end{align}

Since the number of joint trajectories $(\overline{x}_{i},\overline{x}_{j})$ is exponentially large in $T$, an exact representation of the messages is in general computationally unfeasible.
Here, similarly to \citep{barthel2018matrix}, we parametrize messages in terms of matrix product states \cite{fannes1992finitely,verstraete2006matrix,perez2007matrix}, also known as tensor trains in the mathematical literature \cite{oseledets2011tensor}. Following the jargon of tensor networks, in the rest of the paper we will refer to the size of the matrices as bond dimension. 
For a wide class of dynamics including Glauber with $J_{ij}=\pm J$ and epidemic spreading with homogeneous infectivity, the computational cost for a single BP iteration is $\mathcal{O}(T|E|M^6)$ where $T$ is the number of epochs, $|E|$ is the number of edges in the graph and $M$ is the bond dimension.
In all the applications we considered, small bond dimension (scaling at most polynomially with $T$) was enough to obtain almost exact results.  The full description of the approach is found in the Methods section.

\section*{Results}

In this section we illustrate the effectiveness of MPBP applied to
dynamics of epidemic spreading and of the kinetic Ising model. We first focus on free dynamics,
showing results that are at least comparable with the existing methods,
often more accurate. Then we move to reweighted processes, where our
approach really represents an innovation.

\subsection*{Risk assessment in Epidemics}

\begin{figure*}[t]
	\begin{raggedright}
		\includegraphics[width=1\textwidth]{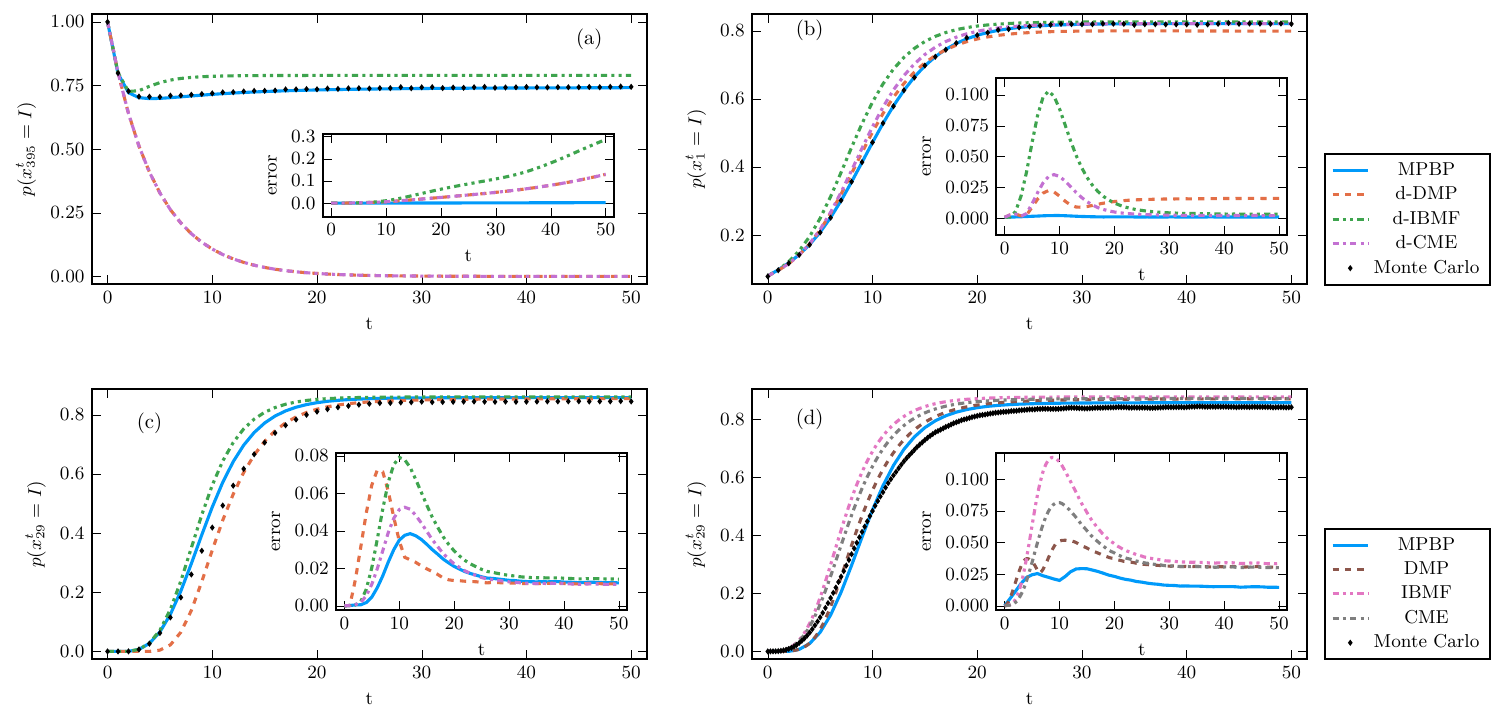}
		\par\end{raggedright}
	\caption{\label{fig:sis_free}Marginal probabilities of free dynamics under
		the SIS model, comparison with models mentioned in the text. The main
		panels correspond to marginals for a single node of the graph, insets
		show the average absolute error over all nodes with respect to Monte
		Carlo simulations. Panels (a-c) compare against discretized versions
		of DMP, IBMF and CME (here with a "d-" prefix) and the Monte Carlo
		strategy reported in the text, panel (d) against regular continuous-time
		versions and a Gillespie-like Monte Carlo simulation. (a) Marginal
		of node $395$, the most connected one of a random tree with $N=1000$
		nodes, $\lambda=0.3,\rho=0.2$. Node $395$ is the only infectious
		at time zero. Bond dimension $12$. (b) Marginal of node $1$
		of a ER graph with $N=1000$ nodes, average connectivity $c=5$, $\lambda=0.1,\rho=0.05,\gamma=0.08$.
		Bond dimension $10$. (c) Marginal of node $29$ (zero-based
		numbering to match previous works) of Zachary's karate club network,
		$N=34$ nodes, $\lambda=0.1,\rho=0.05$, node $0$ is the only infectious
		at time zero. Bond dimension $10$. (d) Same as (c) but the
		comparison is with continuous-time methods, with the addition of CME.}
\end{figure*}

As examples of free dynamics, we estimate the marginal probability
of an individual being in the infectious state under the SIS model,
in several settings (fig. \ref{fig:sis_free}). 
On a random tree and on a diluted random graph,
both of size $N=1000$, MPBP shows almost no discrepancy with Monte
Carlo averages (fig. \ref{fig:sis_free}a, \ref{fig:sis_free}b).
In the former case a single node was picked as the sole infectious
individual at time zero, in the latter a uniform probability $\gamma_{i}\equiv\gamma$
was put on each node. As a comparison we report the curves obtained
using a discrete-time version of Dynamic message Passing (DMP)
\citep{shrestha2015message}, Individual-Based Mean
Field (IBMF) \citep{van2008virus}, and Cavity Master Equation (CME) \cite{aurell2017cavity}, which were originally devised for continuous
time evolution (more details in the supplementary information). We evaluate the accuracy
of each method by considering the average absolute error with respect
to a Monte Carlo simulation $\frac{1}{N}\sum_{i=1}^{N}\left|p(x_{i}^{t}=I)-p^{MC}(x_{i}^{t}=I)\right|$
(insets of fig. \ref{fig:sis_free}). The same analysis is repeated
on Zachary\textquoteright s karate club graph \citep{zachary} (fig.
\ref{fig:sis_free}c), the same benchmark used in \citep{shrestha2015message,ortega2022dynamics}.
It must be pointed out that although MPBP shows by far the best performance
in these comparison, the other considered methods are significantly
simpler. None of the analytic methods is devised to analyze reweighted
dynamics.
Finally, we compare MPBP against three continuous-time methods, DMP,
IBMF and CME, on the karate club graph (fig. \ref{fig:sis_free}d).
The comparison is made by multiplying the transmission and recovery
rates for the continuous setting $\lambda,\rho$ by the time-step
$\Delta t$ (in this case $\Delta t=1$) to turn them into probabilities
to be handled by MPBP. MPBP gives the best overall prediction across
the considered window. 

\begin{figure*}
	\includegraphics[width=1\textwidth]{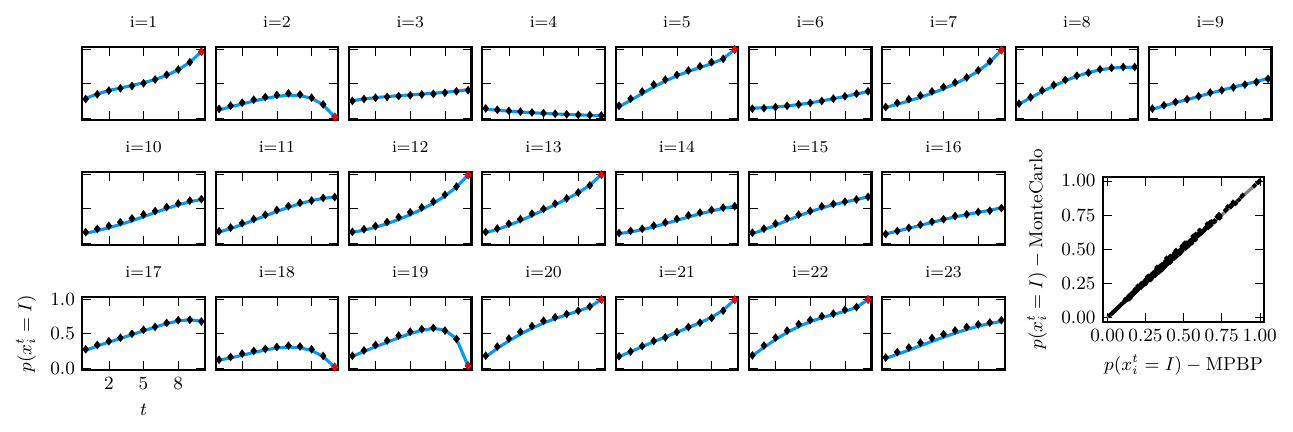}
	\caption{\label{fig:sis_softmargin} MPBP (solid line) with bond dimension $3$ correctly computes marginals of an SIS model defined on an Erdos-Renyi
		graph with $23$ nodes and average connectivity $4$, $\lambda=0.15,\rho=0.12,\gamma=0.13$.
		The state of a random half of the variables was observed at final time $T=10$ and used to reweight the distribution
		(red dots). Black dots are the average over $10^{6}$ Monte Carlo
		simulations. (Bottom-right) Comparison of all points from the previous
		plots, the Pearson correlation coefficient is $0.9986$.}
\end{figure*}

Moving to reweighted processes, fig. \ref{fig:sis_softmargin} shows
the efficacy of MPBP when performing inference of trajectories given
some observations. On a small ($N=23)$ random graph, a 10-step trajectory
$\overline{\boldsymbol{y}}$ was sampled from a SIS prior distribution
with $\lambda=0.15,\rho=0.12,\gamma=0.13$. We then observed the state of a random half $I\subset V$ of the nodes, added the corresponding reweighting factors
$\prod_{i\in I}\phi_{i}^{T}(x_{i}^{T})=\prod_{i\in I}\delta(y_{i}^{T},x_{i}^{T})$
 and performed inference
using \eqref{eq:bayes}. 
The MPBP estimate for the posterior marginals,
obtained with matrices of size $3$, agrees almost perfectly with
Monte Carlo simulations. This is good indication that MPBP applied
to sparse problems will keep giving accurate results even when on
larger and/or more constrained instances where Monte Carlo methods
fail, leaving little to compare against.

\begin{figure*}
	\includegraphics[width=1\textwidth]{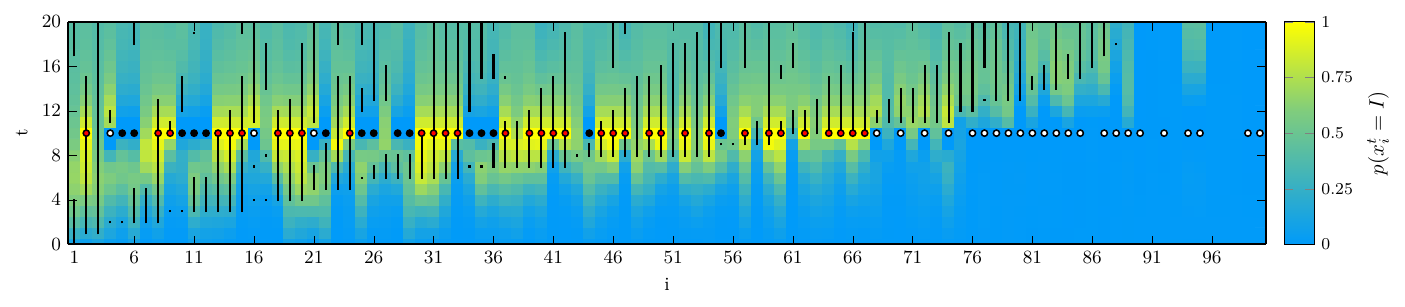}
	\caption{\label{fig:sis_inference_single_instance}Inference on a single epidemic
		outbreak sampled from a SIRS model on an Erdos-Renyi graph with average
		connectivity $c=2.5$, $N=100$. Bond dimension $M=3$. The process to be inferred was drawn
		from a SIRS prior with $\lambda=0.4,\rho=\sigma=0.15,\gamma=0.01$,
		the same parameters were used for the inference. The state of $75\%$ of the nodes
		was observed at time $10$ (white=S, red=I, black=R) and used to reweight the distribution. Black lines correspond to true infection periods.}
\end{figure*}

Realistic scenarios are often better described by the Susceptible-Infectious-Recovered-Susceptible
(SIRS) model where transmission of infections is analogous to the
SIS case, but an infectious node $i$ can recover with probability
$\rho_{i}$ and a recovered become susceptible again with probability
$\sigma_{i}$. From a practical point of view, extending the SIR to
SIRS in the MPBP framework takes little effort: it suffices to enrich
the factors with the new transition $R\to S$. Fig. \ref{fig:sis_inference_single_instance}
shows the performance of MPBP at estimating the posterior trajectories
for a single realization of an epidemic drawn from a prior whose parameters
$\lambda,\rho,\sigma,\gamma$ are homogeneous and known. 
The state of a random
$75\%$ of the system was observed at an intermediate time (colored
dots). We see good agreement between the true infection times (black
lines) and the marginal probabilities of being Infectious (in yellow).
Nodes are sorted in increasing order of true first infection time.

\subsection*{Kinetic Ising}

\begin{figure*}
	\includegraphics[width=1\textwidth]{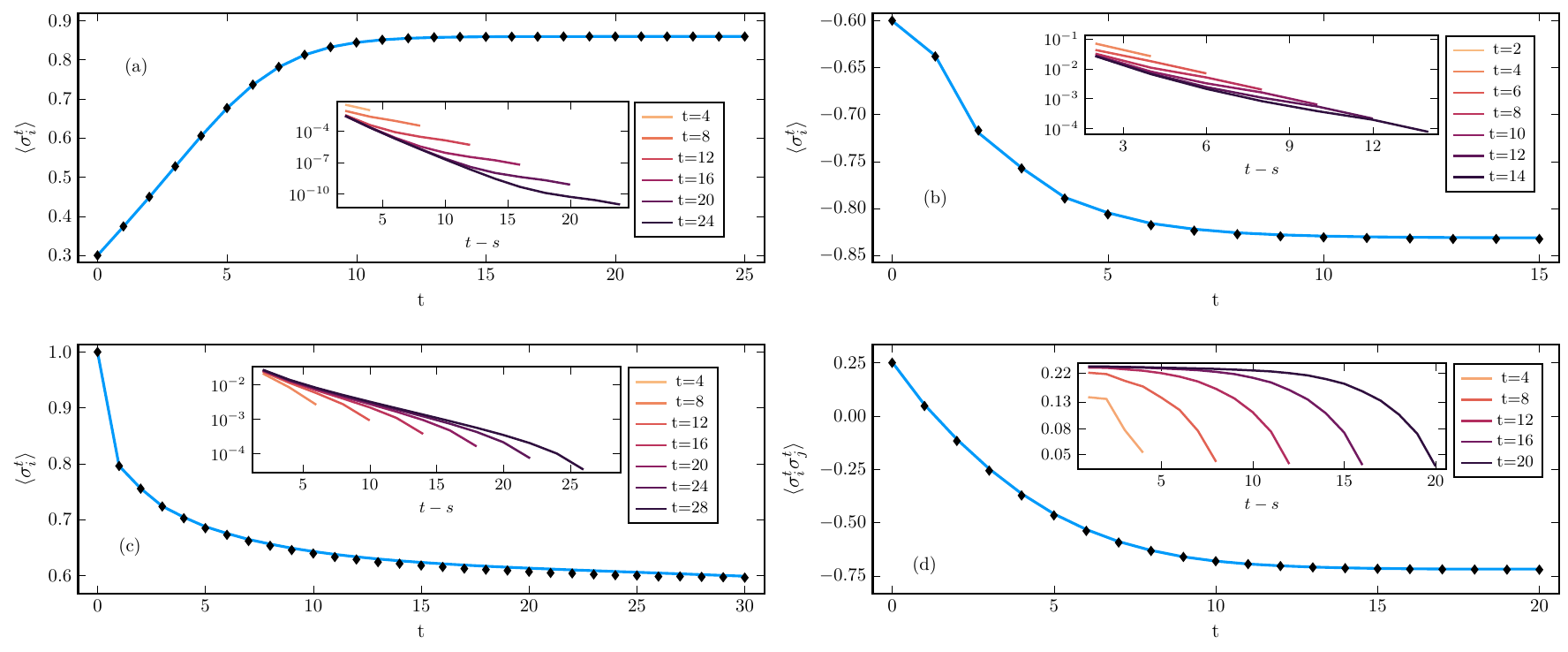}
	\caption{\label{fig:glauber}Magnetization $\langle\sigma_{i}^{t}\rangle$
		(a-c) or nearest-neighbor correlation $\langle\sigma_{i}^{t}\sigma_{j}^{t}\rangle$
		(d) as a function of time for different Ising models. Solid lines
		are MPBP, dots are Monte Carlo simulations on graphs of size $N_{MC}$,
		dashed horizontal lines are the equilibrium values (a-c) or 1RSB prediction
		(d) for the corresponding static versions of the models. Insets show
		autocovariances $\langle\sigma_{i}^{t}\sigma_{i}^{s}\rangle-\langle\sigma_{i}^{t}\rangle\langle\sigma_{i}^{s}\rangle$,
		only even epochs are shown for panels (a-c) because of odd-even effects
		in the dynamics of ferromagnetic models (as in \citep{barthel2018matrix,vazquez2017simple}).
		(a) Infinite $8$-Random Regular Graph, $\beta J=0.2$, $N_{MC}=5000$,
		bond dimension $25$. (b) Infinite Erdos-Renyi graph with mean connectivity
		$c=4$, $\beta J=0.5$, $N_{MC}=5000$, bond dimension $18$. (c) Random
		Field Ising Model on Erdos-Renyi graph with mean connectivity $c=3,\beta J=2/c$,
		$N=N_{MC}=1000$ and $\beta h_{i}=\pm0.6$ sampled uniformly, matrix
		size $10$. (d) Antiferromagnetic Ising Model on infinite $3$-Random
		Regular Graph with $J=-1,\beta=\infty$, $N_{MC}=5000$, bond dimension
		$23$.}
\end{figure*}

As examples of free dynamics we consider the evolution of magnetization
$\langle\sigma_{i}^{t}\rangle$ and time autocovariance $\langle\sigma_{i}^{t}\sigma_{i}^{s}\rangle-\langle\sigma_{i}^{t}\rangle\langle\sigma_{i}^{s}\rangle$
for pairs of epochs $(t,s)$, on ferromagnetic, Random Field and spin-glass
Ising Models (fig. \ref{fig:glauber}), under the stochastic transition \eqref{eq:glauber2}. 
First we consider a model with uniform couplings $J_{ij}\equiv J$ on an infinite Random Regular
Graph like the one studied in \citep{barthel2018matrix} but with
degree $8$ instead of $3$. We then apply our method to an infinite
Erdos-Renyi graph, again with uniform couplings and in the ferromagnetic
phase, using a population dynamics approach. Next, we study a Random
Field Ising Model (RFIM) with uniform couplings and random external
fields $h_{i}=\pm h$ on a large graph. In all three cases the system
is initialized in a magnetized state and the fraction of up spins
grows or decreases monotonically until it reaches a stationary
value. For these second and third models we picked the same settings
as in \citep{vazquez2017simple}. Finally, we consider an antiferromagnetic
model with $J=-1$ at zero temperature ($\beta=\infty$), focusing
on the nearest-neighbor correlation $\langle\sigma_{i}^{t}\sigma_{j}^{t}\rangle$,
$(i,j)\in E$ rather than the magnetization, which is null at steady
state. Above the critical inverse temperature $\beta_{c}=\log\left(1+\sqrt{2}\right)$
\citep{coja2022ising}, the underlying Ising system is in a glassy phase. For this model we used the modified version
of the dynamics reported in (\ref{eq:glauber2}) with $p_{0}=0.25$.

\begin{figure*}
\includegraphics[width=1\textwidth]{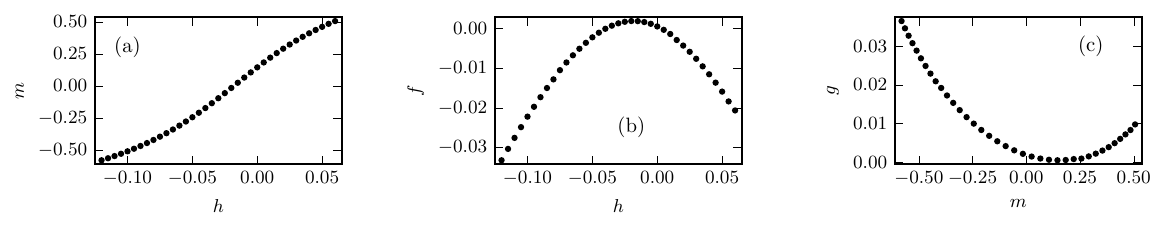}
\caption{\label{fig:large_deviations}Large deviation study of Glauber dynamics
on an infinite 3-Random Regular Graph. Free dynamics with $\beta J=0.6$,
$T=10$, magnetization at time zero $m^{0}=0.1$, zero external field,
reweighted with an external field at final time $\prod_{i}\phi_{i}^{T}(\sigma_{i}^{T})=\prod_{i}e^{h\sigma_{i}^{T}}$.
(a) Magnetization vs reweighting field, (b) Bethe Free Energy vs reweighting
field, (c) Magnetization-constrained free energy $g(m)$ vs magnetization.
Bond dimension $25$.}
\end{figure*}

Finally, we study the large deviation behavior of a free dynamic $W(\overline{\boldsymbol{\sigma}})=\prod_{i=1}^{N}w(\sigma_{i}^{0})\prod_{t=0}^{T-1}w(\sigma_{i}^{t+1}|\boldsymbol{\sigma}_{\partial i}^{t})$
by tilting it with an external field at final time $\prod_{i}\phi_{i}^{T}(\sigma_{i}^{T})=\prod_{i}e^{h\sigma_{i}^{T}}$.
In the thermodynamic limit $N\to\infty$ this allows to select a particular
value for the magnetization at final time $m=\frac{1}{N}\sum_{i}\sigma_{i}^{T}$.
The Bethe Free Energy computed via MPBP is an approximation for
\begin{align}
f(h)= & -\frac{1}{N}\log\sum_{\{\sigma_{i}^{t}\}_{i,t}}W(\overline{\boldsymbol{\sigma}})e^{h\sum_{i}\sigma_{i}^{T}}\\
= & -\frac{1}{N}\log\sum_{m}e^{-N[g(m)-hm]}\\
\xrightarrow{N\to\infty} & \min_{m}\left\{ g(m)-hm\right\} \label{eq:legendre}\\
= & g(m(h))-hm(h)
\end{align}
where $g(m)=-\frac{1}{N}\log\sum_{\{\sigma_{i}^{t}\}_{i,t}}W(\overline{\boldsymbol{\sigma}})\delta\left(Nm,\sum_{i}\sigma_{i}^{T}\right)$,
and $m(h)=\arg\min_{m}\left\{ g(m)-hm\right\} $. In regions where
$g(m)$ is convex, the Legendre transform (\ref{eq:legendre}) can
be inverted to obtain a large deviation law for the probability of
observing the system at final time with magnetization $m$
\begin{equation}
p(m)\sim e^{-N\left[f\left(h(m)\right)+mh(m)\right]}\label{eq:LD}
\end{equation}
where $h(m)$ is the inverse of $m(h)$. Fig. \ref{fig:large_deviations}
shows the estimate of $g(m)$ for a ferromagnetic Ising model on an
infinite random graph initialized at magnetization $m^{0}=0.1$ and
evolving for $T=10$ epochs. $p(m)$
has a minimum at $m\approx0.145$ which corresponds to the free dynamics
$h=0$.

Such an analysis could not have been carried out by means of Monte
Carlo methods since the probability of sampling a trajectory ending
at $m$ is infinitesimal, as is clear from the large deviation law
in fig. \ref{fig:large_deviations}.

\section*{Discussion}

It is often the case that stochastic processes which can be described
accurately, be it by analytical or Monte-Carlo methods, become computationally
difficult as soon as the dynamics is biased by some reweighting factor.
This constitutes a massive limitation since reweighting is essential
whenever one is interested in describing atypical trajectories, an
emblematic example being inference in epidemic models. As of today
there exist, to the best of our knowledge, no analytic method able
to describe reweighted complex dynamics on networks except for the
simple case of non-recurrent models. In this article we adopted the
matrix-product parametrization, inspired by techniques used originally
in quantum physics and recently applied to classical stochastic dynamics
in \citep{barthel2018matrix}, to devise the Matrix Product Belief
Propagation method. We used it to describe reweighted Markov dynamics
on graphs, and applied it to epidemic spreading and a dynamical Ising
models. With respect to the important work in \citep{barthel2018matrix,barthel2020matrix},
which we recall that applies only to free dynamics, our contribution
is twofold. 

First, we develop for MPBP a general scheme to render the computation
time linear in the node degree rather than exponential on a wide class
of models, allowing us to compare it extremely favorably with existing
methods on standard benchmark examples (which typically include vertices
with large degrees). The bottleneck of the whole computation in the
final scheme is due to the SVD factorization, which are cubic in the bond dimension $M$: larger matrices give a better approximation,
but require a greater computational effort. The overall cost per iteration,
assuming the bond dimension constant, is $O\left(T\left|E\right|\right)$,
i.e. linear in the number of edges of the graph. A small number of
iterations is normally sufficient for approximate convergence a fixed
point. A strategy we found to be effective is to start with matrix
size $M$ very small, say $4$ or $5$, iterate until convergence,
then repeat with increasingly larger $M$. It is fair to point out,
however, that although linear, depending on the target accuracy of
the approximation defined by the parameter $M$, the method may be
substantially more computationally intensive than the others used
for comparison.

Second and more importantly, the MPBP approach allows to include reweighting
factors. In particular, the approach proposed in \citep{barthel2018matrix,barthel2020matrix}
is iterated forward in (dynamical) time, and thus allows no backward
flow of information which is necessary with reweighting factors. Reweighting
factors are necessary to analyze conditioned dynamics and rare events.

MPBP, like many other statistical physics-inspired approaches to stochastic
dynamics, is based on the cavity approximation. The Belief Propagation
formalism gives access to the thermodynamic limit for certain ensembles
of random graphs, provides an approximation to the partition function
through the Bethe Free Energy, and allows to compute time autocorrelations.
The limits of validity of MPBP are inherited from those of the cavity
approximation: using the jargon of disordered systems, the approximation
is accurate as long as the problem is in a \textit{Replica Symmetric}
(RS) phase. In the case of epidemic inference presented in fig. \ref{fig:sis_inference_single_instance}
this is surely the case, since the trajectory to be inferred was sampled
from the same prior used for the inference. This amounts to working
on the Nishimori line, where it is known that no replica symmetry-breaking
takes place \citep{iba1999nishimori}. A study of the performance
in regimes where replica symmetry is broken is left for future investigation. 

On graphs with short loops, the performance of BP degrades substantially.
In the static case, this issue can sometimes be overcome by resorting
to higher order approximations \citep{yedidia2000generalized}. We
argue that the same ideas can be translated to dynamics, for example
by describing explicitly the joint trajectory of quadruples of neighboring
variables on a square lattice.

Software implementing the method is available at \citep{repo} and
can be used to directly reproduce the results in the article. The
framework is flexible and accommodates for the inclusion of new models
of dynamics.

As a final remark, we recall that the method applies more in general
to any distribution of the type (\ref{eq:p_general}), where $t$
need not be interpreted as a time index but could, for instance, span
a further spatial direction. Investigation along this line is left
for future work.

% \matmethods{
\section*{Materials and methods}
As anticipated, messages are parametrized in terms of matrix products
\begin{equation}
	\begin{aligned}m_{i\to j}(\overline{x}_{i},\overline{x}_{j})\propto & \prod_{t=0}^{T}A_{i\to j}^{t}(x_{i}^{t},x_{j}^{t})\end{aligned}
	\label{eq:mps}
\end{equation}
where, for any $(x_{i}^{t},x_{j}^{t})$, $A_{i\to j}^{t}(x_{i}^{t},x_{j}^{t})$
is a real-valued matrix. We set $A^{0}$ to have one row and $A^{T}$
to have one column, so that the whole product gives a scalar. Plugging
the \emph{ansatz} (\ref{eq:mps}) into the RHS of the BP equation (\ref{eq:bp}) gives

\begin{equation}
	m_{i\to j}(\overline{x}_{i},\overline{x}_{j})\propto\prod_{t=0}^{T}B_{i\to j}^{t}(x_{i}^{t+1},x_{i}^{t},x_{j}^{t})
\end{equation}
with 
\begin{align}
	B_{i\to j}^{t}(x_{i}^{t+1},x_{i}^{t},x_{j}^{t})= & \sum_{\{x_{k}^{t}\}_{k\in\partial i\setminus j}}f_{i}^{t+1}(x_{i}^{t+1},\boldsymbol{x}_{\partial i}^{t},x_{i}^{t})\nonumber \\
	& \times\left[\bigotimes_{k\in\partial i\setminus j}A_{k\to i}^{t}(x_{k}^{t},x_{i}^{t})\right].\label{eq:exponentialmps}
\end{align}

Two steps are missing in order to close the BP equations under a matrix
product \emph{ansatz}, as discussed in \citep{barthel2018matrix}. First, matrices must be recast into the form (\ref{eq:mps}). Second, if incoming $A$ matrices have bond dimension $M$, $B$ matrices will have bond dimension $M^{|\partial i|-1}$ and
thus will keep growing indefinitely throughout the iterations. Both
issues are solved by means of two successive sweeps of Singular Value
Decompositions (SVD). SVD decomposes a real-valued matrix $A$ as
$A_{ij}=\sum_{k,l=1}^{M}U_{ik}\Lambda_{kl}V_{jl}$ where $\Lambda_{kl}=\lambda_{k}\delta_{k,l}$
is the diagonal matrix of singular values $\lambda_{1}\ge\lambda_{2}\ge\ldots\ge\lambda_{M}\ge0$
and $U^{\dagger}U=VV^{\dagger}=\mathds{1}$ (we use the dagger symbol
for matrix transpose to avoid confusion with the time labels $t,T$,
but all matrices are real-valued). By retaining only the largest $M'$
singular values and setting the others to zero, one can approximate
$A_{ij}$ with $\widetilde{A}_{ij}\coloneqq\sum_{k=1}^{M'}U_{ik}\lambda_{k}V_{jk}$
making an error $\lVert A-\widetilde{A}\rVert_{F}^{2}=\sum_{ij}\left(A_{ij}-\widetilde{A}_{ij}\right)^{2}=\sum_{k=M'+1}^{M}\lambda_{k}^{2}$.
As a result, both $U$ and $V$ are smaller in size.

The first sweep is done from left to right $t=0,1,2,\ldots,T-1$ by performing an SVD decomposition
\begin{equation}
%	\begin{aligned}
		B_{i\to j}^{t}(x_{i}^{t+1},x_{i}^{t},x_{j}^{t})  =C_{i\to j}^{t}(x_{i}^{t},x_{j}^{t})\Lambda^t \left[V^t(x_{i}^{t+1})\right]^{\dagger}
%		B_{i\to j}^{t+1}(x_{i}^{t+2},x_{i}^{t+1},x_{j}^{t+1})  \leftarrow&\Lambda V^{\dagger}(x_{i}^{t+1})\times \\
%		&\times B_{i\to j}^{t+1}(x_{i}^{t+2},x_{i}^{t+1},x_{j}^{t+1})
%	\end{aligned}
	\label{eq:svd-1}
\end{equation}
then redefine $B_{i\to j}^{t+1}(x_{i}^{t+2},x_{i}^{t+1},x_{j}^{t+1})$ as $\Lambda^t \left[V^t(x_{i}^{t+1})\right]^{\dagger}B_{i\to j}^{t+1}(x_{i}^{t+2},x_{i}^{t+1},x_{j}^{t+1})$. The decomposition in \eqref{eq:svd-1} is performed by incorporating
$x_{i}^{t},x_{j}^{t}$ as row indices and $x_{i}^{t+1}$ as column
index (see the supplementary information for more details). At the end of this first sweep, the message looks like
\begin{equation}
	m_{i\to j}(\overline{x}_{i},\overline{x}_{j})=\prod_{t=0}^{T}C_{i\to j}^{t}(x_{i}^{t},x_{j}^{t})
\end{equation}
where, thanks to the properties of the SVD, it holds that
\begin{equation}
	\sum_{x_{i}^{t}x_{j}^{t}}\left[C_{i\to j}^{t}(x_{i}^{t},x_{j}^{t})\right]^{\dagger}C_{i\to j}^{t}(x_{i}^{t},x_{j}^{t})=\mathbb{\mathds{1}}.\label{eq:left_orthogonal}
\end{equation}
%A MPS with such property is said to be in left-canonical form \cite{schollwock2011density}.

At this point the form (\ref{eq:mps}) is recovered: the BP equations
are closed under a matrix product \emph{ansatz}. All that is left to do is perform a second sweep of SVD, this time discarding the smallest
singular values to obtain matrices of reduced size. Going right to
left $t=T,T-1,\ldots,1$, incorporating $(x_{i}^{t},x_{j}^{t})$ as
column indices: 
\begin{equation}
	\begin{aligned}C_{i\to j}^{t}(x_{i}^{t},x_{j}^{t}) & \stackrel{{\rm SVD,trunc}}{\eqqcolon}U^t\Lambda^t A_{i\to j}^{t}(x_{i}^{t},x_{j}^{t})\\
		C_{i\to j}^{t-1}(x_{i}^{t-1},x_{j}^{t-1}) & \leftarrow C_{i\to j}^{t-1}(x_{i}^{t-1},x_{j}^{t-1})U^t\Lambda^t
	\end{aligned}
	\label{eq:svt-1}
\end{equation}

The errors made during the truncations are controlled: consider a
generic step $t$ in the sweep from right to left. The MPS is in the so-called mixed-canonical form \cite{schollwock2011density}:
\begin{equation}
	C^{0}\cdots C^{t}A^{t+1}\cdots A^{T}
\end{equation}
with $C^{0}\cdots C^{t-1}$ left-orthogonal ($C^{\dagger}C=\mathds{1}$)
and $A^{t+1}\cdots A^{T}$ right-orthogonal ($AA^{\dagger}=\mathds{1}$).
$C^{t}$ is neither.

Canonical forms are a useful tool to perform controlled truncations \cite{oseledets2011tensor,schollwock2011density}.
The error in replacing $C^{t}$ by $\widetilde{C}^{t}$
which retains only $M'$ of the $M$ singular values is 
\begin{align}
	\lVert C^{0}\cdots C^{t}A^{t+1}\cdots A^{T}-C^{0}\cdots\widetilde{C}^{t}A^{t+1}\cdots A^{T}\rVert_{F}^{2}\nonumber \\
	=\lVert C^{t}-\widetilde{C}^{t}\rVert_{F}^{2}=\sum_{k=M'+1}^{M}\lambda_{k}^{2}
\end{align}
where the first equality holds thanks to the orthonormality of $C$
and $A$ matrices.
Keeping the MPS in canonical form ensures that the global
error on the matrix product reduces to the local error on $C^{t}$. 

As a side remark, we point out that
there exist techniques to compute directly the SVD truncated to the
$M'$ largest singular values \citep{larsen1998lanczos,baglama2005augmented}.
Such strategies can be advantageous for large $M$ and small
$M'$.

The results in this work were obtained by fixing the number of retained
singular values, and hence the bond dimension. 
Alternatively, given a target threshold $\varepsilon$, one can select $M'$ adaptively such that, e.g. $\frac{\lambda_{M'}}{\sqrt{\sum_{k}\lambda_{k}^{2}}}<\varepsilon$,
as in \citep{barthel2018matrix}.
We find the approach with fixed bond dimension better suited for an iterative solver such as BP, where messages are computed and then overwritten many times before convergence is reached. 
During the first iterations a coarse approximation with small bond dimension is sufficient and helps to keep the computation time under control.
Then, as messages approach a fixed point, one can refine the estimate by either increasing the bond dimension or switching to a threshold-based truncation method.

\subsection*{Bond dimension}
\begin{figure*}[htb]
	\centering
	\includegraphics[width=0.9\textwidth]{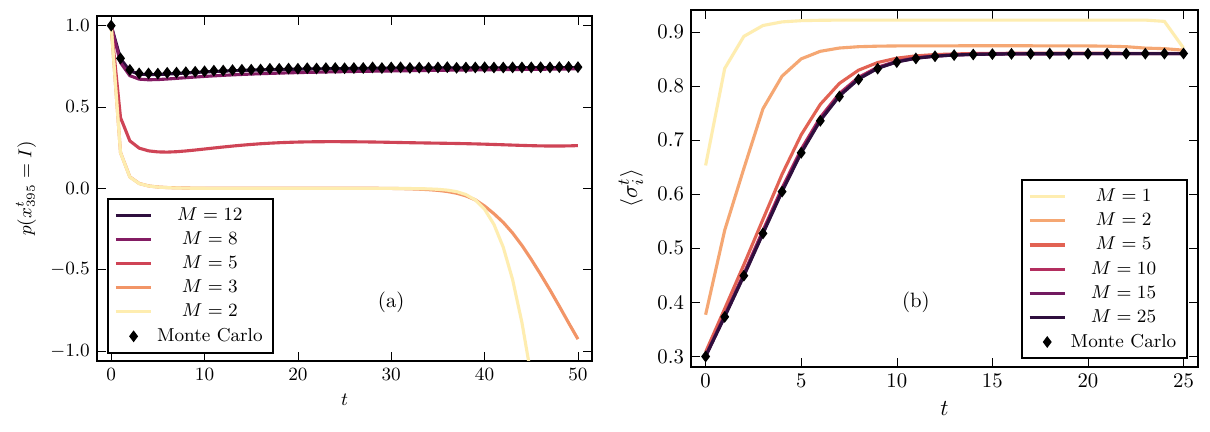}
	\caption{\label{fig:bond_dims} Effect of varying the bond dimension $M$ on the accuracy of the approximation. (a) SIS model on a tree, the same settings as figure \ref{fig:sis_free}a. Too small bond dimension gives unreasonable results. (b) Glauber dynamics on infinite random regular graph of degree $8$, same settings as figure \ref{fig:glauber}a.}
\end{figure*}

Issues may arise whenever excessive truncations turn the matrix product into an ill-defined probability distribution taking negative values.
This is to be expected and indeed was encountered in the experiments we run. Re-running BP with larger bond dimension invariably solved the problem.
Figure \ref{fig:bond_dims} shows the effect of varying the bond dimension in two of the settings shown in the previous plots. 
Instead, truncating too much may lead to unreasonable results such as negative probability values. 

Turning to the expressive power of the MPS \emph{ansatz}, it is reasonable to expect that truncating conservatively, i.e. allowing large bond dimension, will lead to better and better approximations. Indeed, matrix products with arbitrarily large bond dimension can represent exactly any distribution.
However, it is hard to make quantitative statements about the relationship between bond dimension and the complexity that can be captured.  
Based on the discussion in the context of quantum mechanics (see e.g. \cite[section 4.2.2]{schollwock2011density}), it is plausible to assume that strong and long-range (here in time, in the quantum context these are usually in space) correlations need large matrices to be captured accurately. 
However this cannot possibly be the whole story, since there exists a simple counterexample: any trajectory of the SI epidemic model can be represented using MPS of finite bond dimension despite featuring infinite-range correlations. More details are found in the supplementary information.

\subsection*{Convergence}
The BP equations are iterated until convergence to a fixed point.
We opted for an asynchronous update scheme because it tends to feature better convergence properties with respect to a synchronous one. 
Nevertheless, the two can be used interchangeably.
As usual with BP, the procedure naturally lends itself to parallelization, to a larger extent with the synchronous approach.

As a criterion for convergence to a fixed point we computed the marginal distributions at all nodes and epochs $b_i^t(x_i^t)$ (see \eqref{eq:belief}) and checked whether, for an iteration $it$ and the successive one,
\begin{equation}
	\max_{i\in\{1,\ldots,N\}} \max_{t\in\{0,\ldots,T\}}\max_{x_i^t} \left|\left[b_i^t(x_i^t)\right]^{(it+1)}-\left[b_i^t(x_i^t)\right]^{(it)}\right| < \varepsilon
\end{equation}
for some small threshold $\varepsilon$.
A stricter criterion can be considered by computing $\max_{(i,j)\in E}\sum_{\overline{x}_{i},\overline{x}_{j}}\lVert m_{i\to j}(\overline{x}_{i},\overline{x}_{j})^{(it+1)}-m_{i\to j}(\overline{x}_{i},\overline{x}_{j})^{(it)}\rVert_F$. The two criteria lead to similar outcomes (results not shown, see implementation \cite{repo}).

It is worth noting that in the case of free dynamics one can build
the messages incrementally from time $0$ to time $T$ as in DC (see
e.g. \citep{barthel2018matrix}), with no need to iterate until convergence.
Because each sweep of SVD over $t$ matrices takes linear time in
$t$, the total computational cost when using such scheme scales quadratically
with $T$. Instead, initializing messages for all $T$ epochs and
then doing $N_{iter}$ iterations as in our method takes $\mathcal{O}(N_{iter}T)$.
The two are essentially equivalent since we observed that typically
the number of iterations needed to converge is of the order of $T$.
%The procedure naturally lends itself to parallelization: outgoing messages can be computed independently for each node.

It is worth noting that, up to the errors introduced by the truncations,
which we showed to be controlled, MPBP is exact on acyclic graphs.

\subsection*{Observables}

On a fixed point of the BP equations, single-node marginal distributions,  ``beliefs'', are given by
\begin{align}
	b_{i}(\overline{x}_{i})\propto & \sum_{\overline{x}_{\partial i}}\prod_{t=0}^{T-1}f_{i}^{t+1}(x_{i}^{t+1},\boldsymbol{x}_{\partial i}^{t},x_{i}^{t})\nonumber \\
	& \times\prod_{k\in\partial i}m_{k\to i}(\overline{x}_{k},\overline{x}_{i}).\label{eq:belief}
\end{align}

Single-variable and pair marginal distributions as well as time autocorrelations can be computed efficiently on a fixed point of BP by means of standard tensor network contraction techniques (for details, see supplementary information or \cite{oseledets2011tensor}).
The BP formalism also gives access to the Bethe Free Energy, an approximation
to (minus the logarithm of) the normalization of (\ref{eq:p}), which
can be interpreted as the likelihood of the parameters of the dynamics
(e.g. infection rates, temperature,...). In cases where such parameters
are unknown, they can be learned via a maximum-likelihood procedure. 

\subsection*{Thermodynamic limit}

Just like standard BP, MPBP lends itself to be extended to infinite
graphs. In the case of random regular graphs with homogeneous properties
(e.g. $\lambda_{ij}\equiv\lambda,\rho_{i}\equiv\rho$ for epidemic
models, $J_{ij}\equiv J,h_{i}\equiv h$ for Glauber dynamics), a single
message is sufficient to represent the distribution in the thermodynamic
limit. For graph ensembles with variable degree and/or parameters
distributed according to some disorder, we adopt a population dynamics approach (more details in the supplementary information).

\subsection*{A family of models with linear computational cost}

As mentioned before, in the scheme proposed in \citep{barthel2018matrix},
matrices before truncation have size $M^{z-1}$ where $M$ is the
size of matrices in the incoming messages and $z$ is the degree.
The bottleneck are the sweeps of SVDs which yield a computational
cost $\mathcal{O}(M^{3z-3})$ for a single BP message. Although in
a later work \citep[section 6]{barthel2020matrix} it was shown
that such cost can be reduced to $\mathcal{O}(M^{2z-1})$, the exponential
dependence on the degree still represents an issue even for graphs
of moderately large connectivity. Here we show an improved scheme
that, for a wide class of models including many in epidemics and
kinetic Ising, performs the computation in $\mathcal{O}(M^{6})$.
The dependence on $z$ is only polynomial and depends on the details
of the model.

It is enough to notice that in many cases transition probabilities
$w(x_{i}^{t+1}|\boldsymbol{x}_{\partial i}^{t},x_{i}^{t})$ depend
on $\boldsymbol{x}_{\partial i}^{t}$ only through some intermediate
variable which incorporates the aggregate interaction with all the
neighbors. In epidemic models like SI, SIR, SIRS, the transition probability
only depends on the event that at least one of the neighbors has infected
node $i$. In the case of kinetic Ising the transition probability
only depends on the local field, which is a weighted sum of neighboring
spins. 

More formally, consider intermediate scalar variables $y_{A}^{t}$
with $A\subseteq\partial i$ encoding information about $\boldsymbol{x}_{A}^{t}$.
By definition of conditional probability
\begin{align}
	p\left(x_{i}^{t+1}|\boldsymbol{x}_{\partial i}^{t},x_{i}^{t}\right) & =\sum_{y_{\partial i}}p\left(x_{i}^{t+1}|y_{\partial i}^{t},x_{i}^{t}\right)p\left(y_{\partial i}^{t}|\boldsymbol{x}_{\partial i}^{t},x_{i}^{t}\right)
\end{align}
If it holds that 
\begin{align}
	p\left(y_{A\cup B}^{t}|x_{A\cup B}^{t},x_{i}^{t}\right)= & \sum_{y_{A},y_{B}}p\left(y_{A\cup B}^{t}|y_{A}^{t},y_{B}^{t},x_{i}^{t}\right)\nonumber \\
	& \times p\left(y_{A}^{t},y_{B}^{t}|x_{\partial i}^{t},x_{i}^{t}\right)\\
	= & \sum_{y_{A},y_{B}}p\left(y_{A\cup B}^{t}|y_{A}^{t},y_{B}^{t},x_{i}^{t}\right)\nonumber \\
	& \times p\left(y_{A}^{t}|x_{A}^{t},x_{i}^{t}\right)p\left(y_{B}^{t}|x_{B}^{t},x_{i}^{t}\right)\label{eq:pAuB}
\end{align}
for $A\cup B\subseteq\partial i$ (i.e. that the $y$ of disjoint
index sets are independent given the $x$'s), then it suffices to
provide:
\begin{enumerate}
	\item $p\left(y_{j}^{t}|x_{j}^{t},x_{i}^{t}\right)$
	\item $p\left(y_{A\cup B}^{t}|y_{A}^{t},y_{B}^{t},x_{i}^{t}\right)$
\end{enumerate}
to be able to compute the set of outgoing messages from a node in
a recursive manner. This is more efficient than the naive implementation
provided that the number of values that each $y$ can assume does
not grow exponentially with the number of $x$'s it incorporates.
More details of the computation can be found in the supplementary information. 
% }

\begin{acknowledgments}
This study was carried out within the FAIR - Future Artificial Intelligence
Research and received funding from the European Union Next-GenerationEU
(Piano Nazionale di Ripresa e Resilienza (PNRR) -- Missione 4 Componente
2, Investimento 1.3 -- D.D. 1555 11/10/2022, PE00000013). This manuscript
reflects only the authors\textquoteright{} views and opinions, neither
the European Union nor the European Commission can be considered responsible
for them.
\end{acknowledgments}

\bibliographystyle{apsrev4-1}
\bibliography{references}

\begin{thebibliography}{39}%
\makeatletter
\providecommand \@ifxundefined [1]{%
 \@ifx{#1\undefined}
}%
\providecommand \@ifnum [1]{%
 \ifnum #1\expandafter \@firstoftwo
 \else \expandafter \@secondoftwo
 \fi
}%
\providecommand \@ifx [1]{%
 \ifx #1\expandafter \@firstoftwo
 \else \expandafter \@secondoftwo
 \fi
}%
\providecommand \natexlab [1]{#1}%
\providecommand \enquote  [1]{``#1''}%
\providecommand \bibnamefont  [1]{#1}%
\providecommand \bibfnamefont [1]{#1}%
\providecommand \citenamefont [1]{#1}%
\providecommand \href@noop [0]{\@secondoftwo}%
\providecommand \href [0]{\begingroup \@sanitize@url \@href}%
\providecommand \@href[1]{\@@startlink{#1}\@@href}%
\providecommand \@@href[1]{\endgroup#1\@@endlink}%
\providecommand \@sanitize@url [0]{\catcode `\\12\catcode `\$12\catcode
  `\&12\catcode `\#12\catcode `\^12\catcode `\_12\catcode `\%12\relax}%
\providecommand \@@startlink[1]{}%
\providecommand \@@endlink[0]{}%
\providecommand \url  [0]{\begingroup\@sanitize@url \@url }%
\providecommand \@url [1]{\endgroup\@href {#1}{\urlprefix }}%
\providecommand \urlprefix  [0]{URL }%
\providecommand \Eprint [0]{\href }%
\providecommand \doibase [0]{http://dx.doi.org/}%
\providecommand \selectlanguage [0]{\@gobble}%
\providecommand \bibinfo  [0]{\@secondoftwo}%
\providecommand \bibfield  [0]{\@secondoftwo}%
\providecommand \translation [1]{[#1]}%
\providecommand \BibitemOpen [0]{}%
\providecommand \bibitemStop [0]{}%
\providecommand \bibitemNoStop [0]{.\EOS\space}%
\providecommand \EOS [0]{\spacefactor3000\relax}%
\providecommand \BibitemShut  [1]{\csname bibitem#1\endcsname}%
\let\auto@bib@innerbib\@empty
%</preamble>
\bibitem [{\citenamefont {Van~Mieghem}\ \emph {et~al.}(2008)\citenamefont
  {Van~Mieghem}, \citenamefont {Omic},\ and\ \citenamefont
  {Kooij}}]{van2008virus}%
  \BibitemOpen
  \bibfield  {author} {\bibinfo {author} {\bibfnamefont {P.}~\bibnamefont
  {Van~Mieghem}}, \bibinfo {author} {\bibfnamefont {J.}~\bibnamefont {Omic}}, \
  and\ \bibinfo {author} {\bibfnamefont {R.}~\bibnamefont {Kooij}},\
  }\href@noop {} {\bibfield  {journal} {\bibinfo  {journal} {IEEE/ACM
  Transactions On Networking}\ }\textbf {\bibinfo {volume} {17}},\ \bibinfo
  {pages} {1} (\bibinfo {year} {2008})}\BibitemShut {NoStop}%
\bibitem [{\citenamefont {Karrer}\ and\ \citenamefont
  {Newman}(2010)}]{karrer2010message}%
  \BibitemOpen
  \bibfield  {author} {\bibinfo {author} {\bibfnamefont {B.}~\bibnamefont
  {Karrer}}\ and\ \bibinfo {author} {\bibfnamefont {M.~E.}\ \bibnamefont
  {Newman}},\ }\href@noop {} {\bibfield  {journal} {\bibinfo  {journal}
  {Physical Review E}\ }\textbf {\bibinfo {volume} {82}},\ \bibinfo {pages}
  {016101} (\bibinfo {year} {2010})}\BibitemShut {NoStop}%
\bibitem [{\citenamefont {Del~Ferraro}\ and\ \citenamefont
  {Aurell}(2015)}]{delferraro2015dynamic}%
  \BibitemOpen
  \bibfield  {author} {\bibinfo {author} {\bibfnamefont {G.}~\bibnamefont
  {Del~Ferraro}}\ and\ \bibinfo {author} {\bibfnamefont {E.}~\bibnamefont
  {Aurell}},\ }\href {\doibase 10.1103/PhysRevE.92.010102} {\bibfield
  {journal} {\bibinfo  {journal} {Phys. Rev. E}\ }\textbf {\bibinfo {volume}
  {92}},\ \bibinfo {pages} {010102} (\bibinfo {year} {2015})}\BibitemShut
  {NoStop}%
\bibitem [{\citenamefont {Pelizzola}(2013)}]{pelizzola2013variational}%
  \BibitemOpen
  \bibfield  {author} {\bibinfo {author} {\bibfnamefont {A.}~\bibnamefont
  {Pelizzola}},\ }\href@noop {} {\bibfield  {journal} {\bibinfo  {journal} {The
  European Physical Journal B}\ }\textbf {\bibinfo {volume} {86}},\ \bibinfo
  {pages} {1} (\bibinfo {year} {2013})}\BibitemShut {NoStop}%
\bibitem [{\citenamefont {Pelizzola}\ and\ \citenamefont
  {Pretti}(2017)}]{pelizzola2017variational}%
  \BibitemOpen
  \bibfield  {author} {\bibinfo {author} {\bibfnamefont {A.}~\bibnamefont
  {Pelizzola}}\ and\ \bibinfo {author} {\bibfnamefont {M.}~\bibnamefont
  {Pretti}},\ }\href {\doibase 10.1088/1742-5468/aa7a40} {\bibfield  {journal}
  {\bibinfo  {journal} {Journal of Statistical Mechanics: Theory and
  Experiment}\ }\textbf {\bibinfo {volume} {2017}},\ \bibinfo {pages} {073406}
  (\bibinfo {year} {2017})}\BibitemShut {NoStop}%
\bibitem [{\citenamefont {Aurell}\ \emph {et~al.}(2017)\citenamefont {Aurell},
  \citenamefont {Del~Ferraro}, \citenamefont {Dom\'{\i}nguez},\ and\
  \citenamefont {Mulet}}]{aurell2017cavity}%
  \BibitemOpen
  \bibfield  {author} {\bibinfo {author} {\bibfnamefont {E.}~\bibnamefont
  {Aurell}}, \bibinfo {author} {\bibfnamefont {G.}~\bibnamefont {Del~Ferraro}},
  \bibinfo {author} {\bibfnamefont {E.}~\bibnamefont {Dom\'{\i}nguez}}, \ and\
  \bibinfo {author} {\bibfnamefont {R.}~\bibnamefont {Mulet}},\ }\href
  {\doibase 10.1103/PhysRevE.95.052119} {\bibfield  {journal} {\bibinfo
  {journal} {Phys. Rev. E}\ }\textbf {\bibinfo {volume} {95}},\ \bibinfo
  {pages} {052119} (\bibinfo {year} {2017})}\BibitemShut {NoStop}%
\bibitem [{\citenamefont {Ortega}\ \emph {et~al.}(2022)\citenamefont {Ortega},
  \citenamefont {Machado},\ and\ \citenamefont
  {Lage-Castellanos}}]{ortega2022dynamics}%
  \BibitemOpen
  \bibfield  {author} {\bibinfo {author} {\bibfnamefont {E.}~\bibnamefont
  {Ortega}}, \bibinfo {author} {\bibfnamefont {D.}~\bibnamefont {Machado}}, \
  and\ \bibinfo {author} {\bibfnamefont {A.}~\bibnamefont {Lage-Castellanos}},\
  }\href {\doibase 10.1103/PhysRevE.105.024308} {\bibfield  {journal} {\bibinfo
   {journal} {Phys. Rev. E}\ }\textbf {\bibinfo {volume} {105}},\ \bibinfo
  {pages} {024308} (\bibinfo {year} {2022})}\BibitemShut {NoStop}%
\bibitem [{\citenamefont {Neri}\ and\ \citenamefont
  {Boll{\'e}}(2009)}]{neri2009cavity}%
  \BibitemOpen
  \bibfield  {author} {\bibinfo {author} {\bibfnamefont {I.}~\bibnamefont
  {Neri}}\ and\ \bibinfo {author} {\bibfnamefont {D.}~\bibnamefont
  {Boll{\'e}}},\ }\href@noop {} {\bibfield  {journal} {\bibinfo  {journal}
  {Journal of Statistical Mechanics: Theory and Experiment}\ }\textbf {\bibinfo
  {volume} {2009}},\ \bibinfo {pages} {P08009} (\bibinfo {year}
  {2009})}\BibitemShut {NoStop}%
\bibitem [{\citenamefont {Barthel}\ \emph {et~al.}(2018)\citenamefont
  {Barthel}, \citenamefont {De~Bacco},\ and\ \citenamefont
  {Franz}}]{barthel2018matrix}%
  \BibitemOpen
  \bibfield  {author} {\bibinfo {author} {\bibfnamefont {T.}~\bibnamefont
  {Barthel}}, \bibinfo {author} {\bibfnamefont {C.}~\bibnamefont {De~Bacco}}, \
  and\ \bibinfo {author} {\bibfnamefont {S.}~\bibnamefont {Franz}},\
  }\href@noop {} {\bibfield  {journal} {\bibinfo  {journal} {Physical Review
  E}\ }\textbf {\bibinfo {volume} {97}},\ \bibinfo {pages} {010104} (\bibinfo
  {year} {2018})}\BibitemShut {NoStop}%
\bibitem [{\citenamefont {Barthel}(2020)}]{barthel2020matrix}%
  \BibitemOpen
  \bibfield  {author} {\bibinfo {author} {\bibfnamefont {T.}~\bibnamefont
  {Barthel}},\ }\href@noop {} {\bibfield  {journal} {\bibinfo  {journal}
  {Journal of Statistical Mechanics: Theory and Experiment}\ }\textbf {\bibinfo
  {volume} {2020}},\ \bibinfo {pages} {013217} (\bibinfo {year}
  {2020})}\BibitemShut {NoStop}%
\bibitem [{\citenamefont {Perez-Garcia}\ \emph {et~al.}(2007)\citenamefont
  {Perez-Garcia}, \citenamefont {Verstraete}, \citenamefont {Wolf},\ and\
  \citenamefont {Cirac}}]{perez2007matrix}%
  \BibitemOpen
  \bibfield  {author} {\bibinfo {author} {\bibfnamefont {D.}~\bibnamefont
  {Perez-Garcia}}, \bibinfo {author} {\bibfnamefont {F.}~\bibnamefont
  {Verstraete}}, \bibinfo {author} {\bibfnamefont {M.~M.}\ \bibnamefont
  {Wolf}}, \ and\ \bibinfo {author} {\bibfnamefont {J.~I.}\ \bibnamefont
  {Cirac}},\ }\href@noop {} {\bibfield  {journal} {\bibinfo  {journal} {Quantum
  Info. Comput.}\ }\textbf {\bibinfo {volume} {7}},\ \bibinfo {pages} {401}
  (\bibinfo {year} {2007})}\BibitemShut {NoStop}%
\bibitem [{\citenamefont {Verstraete}\ and\ \citenamefont
  {Cirac}(2006)}]{verstraete2006matrix}%
  \BibitemOpen
  \bibfield  {author} {\bibinfo {author} {\bibfnamefont {F.}~\bibnamefont
  {Verstraete}}\ and\ \bibinfo {author} {\bibfnamefont {J.~I.}\ \bibnamefont
  {Cirac}},\ }\href@noop {} {\bibfield  {journal} {\bibinfo  {journal}
  {Physical review b}\ }\textbf {\bibinfo {volume} {73}},\ \bibinfo {pages}
  {094423} (\bibinfo {year} {2006})}\BibitemShut {NoStop}%
\bibitem [{\citenamefont {Fannes}\ \emph {et~al.}(1992)\citenamefont {Fannes},
  \citenamefont {Nachtergaele},\ and\ \citenamefont
  {Werner}}]{fannes1992finitely}%
  \BibitemOpen
  \bibfield  {author} {\bibinfo {author} {\bibfnamefont {M.}~\bibnamefont
  {Fannes}}, \bibinfo {author} {\bibfnamefont {B.}~\bibnamefont
  {Nachtergaele}}, \ and\ \bibinfo {author} {\bibfnamefont {R.~F.}\
  \bibnamefont {Werner}},\ }\href@noop {} {\bibfield  {journal} {\bibinfo
  {journal} {Communications in mathematical physics}\ }\textbf {\bibinfo
  {volume} {144}},\ \bibinfo {pages} {443} (\bibinfo {year}
  {1992})}\BibitemShut {NoStop}%
\bibitem [{\citenamefont {Derrida}\ \emph {et~al.}(1993)\citenamefont
  {Derrida}, \citenamefont {Evans}, \citenamefont {Hakim},\ and\ \citenamefont
  {Pasquier}}]{derrida1993exact}%
  \BibitemOpen
  \bibfield  {author} {\bibinfo {author} {\bibfnamefont {B.}~\bibnamefont
  {Derrida}}, \bibinfo {author} {\bibfnamefont {M.~R.}\ \bibnamefont {Evans}},
  \bibinfo {author} {\bibfnamefont {V.}~\bibnamefont {Hakim}}, \ and\ \bibinfo
  {author} {\bibfnamefont {V.}~\bibnamefont {Pasquier}},\ }\href@noop {}
  {\bibfield  {journal} {\bibinfo  {journal} {Journal of Physics A:
  Mathematical and General}\ }\textbf {\bibinfo {volume} {26}},\ \bibinfo
  {pages} {1493} (\bibinfo {year} {1993})}\BibitemShut {NoStop}%
\bibitem [{\citenamefont {Ba{\~n}uls}\ and\ \citenamefont
  {Garrahan}(2019)}]{banuls2019using}%
  \BibitemOpen
  \bibfield  {author} {\bibinfo {author} {\bibfnamefont {M.~C.}\ \bibnamefont
  {Ba{\~n}uls}}\ and\ \bibinfo {author} {\bibfnamefont {J.~P.}\ \bibnamefont
  {Garrahan}},\ }\href@noop {} {\bibfield  {journal} {\bibinfo  {journal}
  {Physical review letters}\ }\textbf {\bibinfo {volume} {123}},\ \bibinfo
  {pages} {200601} (\bibinfo {year} {2019})}\BibitemShut {NoStop}%
\bibitem [{\citenamefont {Han}\ \emph {et~al.}(2018)\citenamefont {Han},
  \citenamefont {Wang}, \citenamefont {Fan}, \citenamefont {Wang},\ and\
  \citenamefont {Zhang}}]{han2018unsupervised}%
  \BibitemOpen
  \bibfield  {author} {\bibinfo {author} {\bibfnamefont {Z.-Y.}\ \bibnamefont
  {Han}}, \bibinfo {author} {\bibfnamefont {J.}~\bibnamefont {Wang}}, \bibinfo
  {author} {\bibfnamefont {H.}~\bibnamefont {Fan}}, \bibinfo {author}
  {\bibfnamefont {L.}~\bibnamefont {Wang}}, \ and\ \bibinfo {author}
  {\bibfnamefont {P.}~\bibnamefont {Zhang}},\ }\href@noop {} {\bibfield
  {journal} {\bibinfo  {journal} {Physical Review X}\ }\textbf {\bibinfo
  {volume} {8}},\ \bibinfo {pages} {031012} (\bibinfo {year}
  {2018})}\BibitemShut {NoStop}%
\bibitem [{\citenamefont {Stoudenmire}\ and\ \citenamefont
  {Schwab}(2016)}]{stoudenmire2016supervised}%
  \BibitemOpen
  \bibfield  {author} {\bibinfo {author} {\bibfnamefont {E.}~\bibnamefont
  {Stoudenmire}}\ and\ \bibinfo {author} {\bibfnamefont {D.~J.}\ \bibnamefont
  {Schwab}},\ }\href@noop {} {\bibfield  {journal} {\bibinfo  {journal}
  {Advances in neural information processing systems}\ }\textbf {\bibinfo
  {volume} {29}} (\bibinfo {year} {2016})}\BibitemShut {NoStop}%
\bibitem [{\citenamefont {Altarelli}\ \emph {et~al.}(2013)\citenamefont
  {Altarelli}, \citenamefont {Braunstein}, \citenamefont {Dall'Asta},\ and\
  \citenamefont {Zecchina}}]{altarelli_large_2013}%
  \BibitemOpen
  \bibfield  {author} {\bibinfo {author} {\bibfnamefont {F.}~\bibnamefont
  {Altarelli}}, \bibinfo {author} {\bibfnamefont {A.}~\bibnamefont
  {Braunstein}}, \bibinfo {author} {\bibfnamefont {L.}~\bibnamefont
  {Dall'Asta}}, \ and\ \bibinfo {author} {\bibfnamefont {R.}~\bibnamefont
  {Zecchina}},\ }\href {\doibase 10.1103/PhysRevE.87.062115} {\bibfield
  {journal} {\bibinfo  {journal} {Physical Review E}\ }\textbf {\bibinfo
  {volume} {87}},\ \bibinfo {pages} {062115} (\bibinfo {year}
  {2013})}\BibitemShut {NoStop}%
\bibitem [{\citenamefont {Crotti}\ and\ \citenamefont
  {Braunstein}(2023)}]{repo}%
  \BibitemOpen
  \bibfield  {author} {\bibinfo {author} {\bibfnamefont {S.}~\bibnamefont
  {Crotti}}\ and\ \bibinfo {author} {\bibfnamefont {A.}~\bibnamefont
  {Braunstein}},\ }\href@noop {} {\enquote {\bibinfo {title}
  {Matrix{P}roduct{B}{P}},}\ }\bibinfo {howpublished}
  {\url{https://github.com/stecrotti/MatrixProductBP.jl}} (\bibinfo {year}
  {2023}),\ \bibinfo {note} {accessed 27/10/2023}\BibitemShut {NoStop}%
\bibitem [{\citenamefont {Renart}\ \emph {et~al.}(2010)\citenamefont {Renart},
  \citenamefont {De~La~Rocha}, \citenamefont {Bartho}, \citenamefont
  {Hollender}, \citenamefont {Parga}, \citenamefont {Reyes},\ and\
  \citenamefont {Harris}}]{renart2010asynchronous}%
  \BibitemOpen
  \bibfield  {author} {\bibinfo {author} {\bibfnamefont {A.}~\bibnamefont
  {Renart}}, \bibinfo {author} {\bibfnamefont {J.}~\bibnamefont {De~La~Rocha}},
  \bibinfo {author} {\bibfnamefont {P.}~\bibnamefont {Bartho}}, \bibinfo
  {author} {\bibfnamefont {L.}~\bibnamefont {Hollender}}, \bibinfo {author}
  {\bibfnamefont {N.}~\bibnamefont {Parga}}, \bibinfo {author} {\bibfnamefont
  {A.}~\bibnamefont {Reyes}}, \ and\ \bibinfo {author} {\bibfnamefont {K.~D.}\
  \bibnamefont {Harris}},\ }\href@noop {} {\bibfield  {journal} {\bibinfo
  {journal} {science}\ }\textbf {\bibinfo {volume} {327}},\ \bibinfo {pages}
  {587} (\bibinfo {year} {2010})}\BibitemShut {NoStop}%
\bibitem [{\citenamefont {Roudi}\ and\ \citenamefont
  {Hertz}(2011)}]{roudi2011mean}%
  \BibitemOpen
  \bibfield  {author} {\bibinfo {author} {\bibfnamefont {Y.}~\bibnamefont
  {Roudi}}\ and\ \bibinfo {author} {\bibfnamefont {J.}~\bibnamefont {Hertz}},\
  }\href@noop {} {\bibfield  {journal} {\bibinfo  {journal} {Physical review
  letters}\ }\textbf {\bibinfo {volume} {106}},\ \bibinfo {pages} {048702}
  (\bibinfo {year} {2011})}\BibitemShut {NoStop}%
\bibitem [{\citenamefont {Ohta}\ and\ \citenamefont
  {Sasa}(2010)}]{ohta2010universal}%
  \BibitemOpen
  \bibfield  {author} {\bibinfo {author} {\bibfnamefont {H.}~\bibnamefont
  {Ohta}}\ and\ \bibinfo {author} {\bibfnamefont {S.-i.}\ \bibnamefont
  {Sasa}},\ }\href@noop {} {\bibfield  {journal} {\bibinfo  {journal}
  {Europhysics Letters}\ }\textbf {\bibinfo {volume} {90}},\ \bibinfo {pages}
  {27008} (\bibinfo {year} {2010})}\BibitemShut {NoStop}%
\bibitem [{\citenamefont {Van~Mieghem}(2011)}]{van2011n}%
  \BibitemOpen
  \bibfield  {author} {\bibinfo {author} {\bibfnamefont {P.}~\bibnamefont
  {Van~Mieghem}},\ }\href@noop {} {\bibfield  {journal} {\bibinfo  {journal}
  {Computing}\ }\textbf {\bibinfo {volume} {93}},\ \bibinfo {pages} {147}
  (\bibinfo {year} {2011})}\BibitemShut {NoStop}%
\bibitem [{\citenamefont {Shrestha}\ \emph {et~al.}(2015)\citenamefont
  {Shrestha}, \citenamefont {Scarpino},\ and\ \citenamefont
  {Moore}}]{shrestha2015message}%
  \BibitemOpen
  \bibfield  {author} {\bibinfo {author} {\bibfnamefont {M.}~\bibnamefont
  {Shrestha}}, \bibinfo {author} {\bibfnamefont {S.~V.}\ \bibnamefont
  {Scarpino}}, \ and\ \bibinfo {author} {\bibfnamefont {C.}~\bibnamefont
  {Moore}},\ }\href@noop {} {\bibfield  {journal} {\bibinfo  {journal}
  {Physical Review E}\ }\textbf {\bibinfo {volume} {92}},\ \bibinfo {pages}
  {022821} (\bibinfo {year} {2015})}\BibitemShut {NoStop}%
\bibitem [{\citenamefont {V{\'a}zquez}\ \emph {et~al.}(2017)\citenamefont
  {V{\'a}zquez}, \citenamefont {Del~Ferraro},\ and\ \citenamefont
  {Ricci-Tersenghi}}]{vazquez2017simple}%
  \BibitemOpen
  \bibfield  {author} {\bibinfo {author} {\bibfnamefont {E.~D.}\ \bibnamefont
  {V{\'a}zquez}}, \bibinfo {author} {\bibfnamefont {G.}~\bibnamefont
  {Del~Ferraro}}, \ and\ \bibinfo {author} {\bibfnamefont {F.}~\bibnamefont
  {Ricci-Tersenghi}},\ }\href@noop {} {\bibfield  {journal} {\bibinfo
  {journal} {Journal of Statistical Mechanics: Theory and Experiment}\ }\textbf
  {\bibinfo {volume} {2017}},\ \bibinfo {pages} {033303} (\bibinfo {year}
  {2017})}\BibitemShut {NoStop}%
\bibitem [{\citenamefont {Del~Ferraro}\ and\ \citenamefont
  {Aurell}(2014)}]{del2014perturbative}%
  \BibitemOpen
  \bibfield  {author} {\bibinfo {author} {\bibfnamefont {G.}~\bibnamefont
  {Del~Ferraro}}\ and\ \bibinfo {author} {\bibfnamefont {E.}~\bibnamefont
  {Aurell}},\ }\href@noop {} {\bibfield  {journal} {\bibinfo  {journal}
  {Journal of the Physical Society of Japan}\ }\textbf {\bibinfo {volume}
  {83}},\ \bibinfo {pages} {084001} (\bibinfo {year} {2014})}\BibitemShut
  {NoStop}%
\bibitem [{\citenamefont {Antulov-Fantulin}\ \emph {et~al.}(2015)\citenamefont
  {Antulov-Fantulin}, \citenamefont {Lan{\v{c}}i{\'c}}, \citenamefont
  {{\v{S}}muc}, \citenamefont {{\v{S}}tefan{\v{c}}i{\'c}},\ and\ \citenamefont
  {{\v{S}}iki{\'c}}}]{antulov2015identification}%
  \BibitemOpen
  \bibfield  {author} {\bibinfo {author} {\bibfnamefont {N.}~\bibnamefont
  {Antulov-Fantulin}}, \bibinfo {author} {\bibfnamefont {A.}~\bibnamefont
  {Lan{\v{c}}i{\'c}}}, \bibinfo {author} {\bibfnamefont {T.}~\bibnamefont
  {{\v{S}}muc}}, \bibinfo {author} {\bibfnamefont {H.}~\bibnamefont
  {{\v{S}}tefan{\v{c}}i{\'c}}}, \ and\ \bibinfo {author} {\bibfnamefont
  {M.}~\bibnamefont {{\v{S}}iki{\'c}}},\ }\href@noop {} {\bibfield  {journal}
  {\bibinfo  {journal} {Physical review letters}\ }\textbf {\bibinfo {volume}
  {114}},\ \bibinfo {pages} {248701} (\bibinfo {year} {2015})}\BibitemShut
  {NoStop}%
\bibitem [{\citenamefont {Oseledets}(2011)}]{oseledets2011tensor}%
  \BibitemOpen
  \bibfield  {author} {\bibinfo {author} {\bibfnamefont {I.~V.}\ \bibnamefont
  {Oseledets}},\ }\href@noop {} {\bibfield  {journal} {\bibinfo  {journal}
  {SIAM Journal on Scientific Computing}\ }\textbf {\bibinfo {volume} {33}},\
  \bibinfo {pages} {2295} (\bibinfo {year} {2011})}\BibitemShut {NoStop}%
\bibitem [{\citenamefont {Kunegis}()}]{zachary}%
  \BibitemOpen
  \bibfield  {author} {\bibinfo {author} {\bibfnamefont {J.}~\bibnamefont
  {Kunegis}},\ }\href@noop {} {\enquote {\bibinfo {title} {Zachary karate
  club},}\ }\bibinfo {howpublished}
  {\url{http://konect.cc/networks/ucidata-zachary/}},\ \bibinfo {note}
  {accessed: 14.01.2023}\BibitemShut {NoStop}%
\bibitem [{\citenamefont {Coja-Oghlan}\ \emph {et~al.}(2022)\citenamefont
  {Coja-Oghlan}, \citenamefont {Loick}, \citenamefont {Mezei},\ and\
  \citenamefont {Sorkin}}]{coja2022ising}%
  \BibitemOpen
  \bibfield  {author} {\bibinfo {author} {\bibfnamefont {A.}~\bibnamefont
  {Coja-Oghlan}}, \bibinfo {author} {\bibfnamefont {P.}~\bibnamefont {Loick}},
  \bibinfo {author} {\bibfnamefont {B.~F.}\ \bibnamefont {Mezei}}, \ and\
  \bibinfo {author} {\bibfnamefont {G.~B.}\ \bibnamefont {Sorkin}},\
  }\href@noop {} {\bibfield  {journal} {\bibinfo  {journal} {SIAM Journal on
  Discrete Mathematics}\ }\textbf {\bibinfo {volume} {36}},\ \bibinfo {pages}
  {1306} (\bibinfo {year} {2022})}\BibitemShut {NoStop}%
\bibitem [{\citenamefont {Iba}(1999)}]{iba1999nishimori}%
  \BibitemOpen
  \bibfield  {author} {\bibinfo {author} {\bibfnamefont {Y.}~\bibnamefont
  {Iba}},\ }\href@noop {} {\bibfield  {journal} {\bibinfo  {journal} {Journal
  of Physics A: Mathematical and General}\ }\textbf {\bibinfo {volume} {32}},\
  \bibinfo {pages} {3875} (\bibinfo {year} {1999})}\BibitemShut {NoStop}%
\bibitem [{\citenamefont {Yedidia}\ \emph {et~al.}(2000)\citenamefont
  {Yedidia}, \citenamefont {Freeman},\ and\ \citenamefont
  {Weiss}}]{yedidia2000generalized}%
  \BibitemOpen
  \bibfield  {author} {\bibinfo {author} {\bibfnamefont {J.~S.}\ \bibnamefont
  {Yedidia}}, \bibinfo {author} {\bibfnamefont {W.}~\bibnamefont {Freeman}}, \
  and\ \bibinfo {author} {\bibfnamefont {Y.}~\bibnamefont {Weiss}},\
  }\href@noop {} {\bibfield  {journal} {\bibinfo  {journal} {Advances in neural
  information processing systems}\ }\textbf {\bibinfo {volume} {13}} (\bibinfo
  {year} {2000})}\BibitemShut {NoStop}%
\bibitem [{\citenamefont {Schollw{\"o}ck}(2011)}]{schollwock2011density}%
  \BibitemOpen
  \bibfield  {author} {\bibinfo {author} {\bibfnamefont {U.}~\bibnamefont
  {Schollw{\"o}ck}},\ }\href@noop {} {\bibfield  {journal} {\bibinfo  {journal}
  {Annals of physics}\ }\textbf {\bibinfo {volume} {326}},\ \bibinfo {pages}
  {96} (\bibinfo {year} {2011})}\BibitemShut {NoStop}%
\bibitem [{\citenamefont {Larsen}(1998)}]{larsen1998lanczos}%
  \BibitemOpen
  \bibfield  {author} {\bibinfo {author} {\bibfnamefont {R.~M.}\ \bibnamefont
  {Larsen}},\ }\href@noop {} {\bibfield  {journal} {\bibinfo  {journal} {DAIMI
  Report Series}\ } (\bibinfo {year} {1998})}\BibitemShut {NoStop}%
\bibitem [{\citenamefont {Baglama}\ and\ \citenamefont
  {Reichel}(2005)}]{baglama2005augmented}%
  \BibitemOpen
  \bibfield  {author} {\bibinfo {author} {\bibfnamefont {J.}~\bibnamefont
  {Baglama}}\ and\ \bibinfo {author} {\bibfnamefont {L.}~\bibnamefont
  {Reichel}},\ }\href@noop {} {\bibfield  {journal} {\bibinfo  {journal} {SIAM
  Journal on Scientific Computing}\ }\textbf {\bibinfo {volume} {27}},\
  \bibinfo {pages} {19} (\bibinfo {year} {2005})}\BibitemShut {NoStop}%
\bibitem [{\citenamefont {Glauber}(1963)}]{glauber1963time}%
  \BibitemOpen
  \bibfield  {author} {\bibinfo {author} {\bibfnamefont {R.~J.}\ \bibnamefont
  {Glauber}},\ }\href@noop {} {\bibfield  {journal} {\bibinfo  {journal}
  {Journal of mathematical physics}\ }\textbf {\bibinfo {volume} {4}},\
  \bibinfo {pages} {294} (\bibinfo {year} {1963})}\BibitemShut {NoStop}%
\bibitem [{\citenamefont {Peretto}(1984)}]{peretto1984collective}%
  \BibitemOpen
  \bibfield  {author} {\bibinfo {author} {\bibfnamefont {P.}~\bibnamefont
  {Peretto}},\ }\href@noop {} {\bibfield  {journal} {\bibinfo  {journal}
  {Biological cybernetics}\ }\textbf {\bibinfo {volume} {50}},\ \bibinfo
  {pages} {51} (\bibinfo {year} {1984})}\BibitemShut {NoStop}%
\bibitem [{\citenamefont {Mezard}\ and\ \citenamefont
  {Montanari}(2009)}]{mezard2009information}%
  \BibitemOpen
  \bibfield  {author} {\bibinfo {author} {\bibfnamefont {M.}~\bibnamefont
  {Mezard}}\ and\ \bibinfo {author} {\bibfnamefont {A.}~\bibnamefont
  {Montanari}},\ }\href@noop {} {\emph {\bibinfo {title} {Information, physics,
  and computation}}}\ (\bibinfo  {publisher} {Oxford University Press},\
  \bibinfo {year} {2009})\BibitemShut {NoStop}%
\bibitem [{\citenamefont {Altarelli}\ \emph {et~al.}(2014)\citenamefont
  {Altarelli}, \citenamefont {Braunstein}, \citenamefont {Dall'Asta},
  \citenamefont {Lage-Castellanos},\ and\ \citenamefont
  {Zecchina}}]{altarelli2014bayesian}%
  \BibitemOpen
  \bibfield  {author} {\bibinfo {author} {\bibfnamefont {F.}~\bibnamefont
  {Altarelli}}, \bibinfo {author} {\bibfnamefont {A.}~\bibnamefont
  {Braunstein}}, \bibinfo {author} {\bibfnamefont {L.}~\bibnamefont
  {Dall'Asta}}, \bibinfo {author} {\bibfnamefont {A.}~\bibnamefont
  {Lage-Castellanos}}, \ and\ \bibinfo {author} {\bibfnamefont
  {R.}~\bibnamefont {Zecchina}},\ }\href@noop {} {\bibfield  {journal}
  {\bibinfo  {journal} {Physical review letters}\ }\textbf {\bibinfo {volume}
  {112}},\ \bibinfo {pages} {118701} (\bibinfo {year} {2014})}\BibitemShut
  {NoStop}%
\end{thebibliography}%

%\appendix
%\documentclass[english,aps,amsmath,amssymb,pre]{revtex4-2}
%
%\usepackage{xr}
%\externaldocument{matrixbp}
%\setcounter{section}{0}
%\renewcommand{\thesection}{S-\arabic{section}}
%\def\theequation{S\arabic{equation}}
%\newcommand{\x}{\boldsymbol{x}}
%\newcommand{\ox}{\overline x}
%\newcommand{\oy}{\overline y}
%\newcommand{\os}{\overline \sigma}
%\newcommand*{\myprop}{\mathrel{\propto}}
%\DeclareMathOperator{\sign}{sign}
%\usepackage{dsfont} % for double-stroke `1`
%\usepackage{amsmath}
%\usepackage{mathtools}
%
%\usepackage{algorithm}
%\usepackage{algpseudocode}
%
%\title{Matrix Product Belief Propagation for reweighted stochastic dynamics over graphs}
%
%\begin{document}
%	
%\maketitle
%\tableofcontents

%%%%%%%%%% Merge with supplemental materials %%%%%%%%%%
\clearpage{}
\onecolumngrid
\begin{center}
	\textbf{\large Supplementary information}
\end{center}
%%%%%%%%%% Merge with supplemental materials %%%%%%%%%%
%%%%%%%%%% Prefix a "S" to all equations, figures, tables and reset the counter %%%%%%%%%%
\setcounter{equation}{0}
\setcounter{figure}{0}
\setcounter{table}{0}
\setcounter{page}{1}
\makeatletter
\renewcommand{\theequation}{S\arabic{equation}}
\renewcommand{\thefigure}{S\arabic{figure}}

\section{Parallel Glauber dynamics and equilibrium}
\subsection{Marginals and correlations in Parallel Glauber dynamics}
It is well known that (fully symmetric) Glauber dynamics with asynchronous update converges to the equilibrium distribution for the underlying Ising model on a graph $G=(V,E)$ \cite{glauber1963time} (a fact that can be trivially verified by checking the detailed balance condition)
\begin{equation}\label{eq:p_ising_eq}
	p^{eq}(\boldsymbol{\sigma})\propto\exp\left\{\beta\left[\sum_{(ij)\in E}J_{ij}\sigma_i\sigma_j+\sum_{i=1}^N h_i\sigma_i\right]\right\}.  
\end{equation}

Parallel updates like the ones considered in this work, instead, lead to a stationary distribution \cite{peretto1984collective}
\begin{equation}\label{eq:p_ising_stat}
	p^{stat}(\boldsymbol{\sigma})\propto\exp\left\{\sum_i \left[ \log\cosh\beta\left(\sum_{j\in\partial i} J_{ij}\sigma_j + h_i\right)+\beta h_i\sigma_i\right]\right\}.  
\end{equation}

Here we show that, provided that the underlying model lives on a bipartite graph:
\begin{enumerate}
	\item The two distributions have the same marginals, i.e. $p^{eq}(\sigma_i)=p^{stat}(\sigma_i)$.
	\item The joint distribution for neighboring variables $p^{eq}(\sigma_i,\sigma_j)$ is equal to $p(\sigma_i^{t+1},\sigma_j^t)$ where $\boldsymbol{\sigma}^t,\boldsymbol{\sigma}^{t+1}$ are configurations sampled using the parallel Glauber update at the stationary state.
\end{enumerate} 

To see why the two propositions are true, consider an augmented system $\tilde{G}=(\tilde V,\tilde E)$ consisting of two copies of the vertices of the original graph. The new system is made of $2N$ variables $\{\sigma_1,\ldots,\sigma_N,\sigma'_1,\ldots,\sigma'_N\}$. Each variable $\sigma_i$ interacts with the copies of its neighbors in the original graph $\{\sigma'_j\}_{j:(ij)\in E}$, and vice-versa. The new system is distributed according to
\begin{equation}
	p^{aug}(\boldsymbol{\sigma},\boldsymbol{\sigma'})\propto \exp\left\{\beta \sum_{i=1}^N\left[\sum_{j:(ij)\in E}J_{ij}\sigma_i\sigma'_j+\sum_i h_i(\sigma_i+\sigma'_i)\right]\right\}.
\end{equation}

By marginalizing over $\boldsymbol{\sigma}$ or $\boldsymbol{\sigma'}$, it is easy to see that either subset is distributed according to $p^{stat}$.
Moreover, because the original graph $G$ was bipartite ($V=A\cup B, A\cap B=\emptyset$), the new graph $\tilde G$ is made of two disconnected components: the first contains variables $\{\sigma_i\}_{i\in A} \cup \{\sigma'_j\}_{j\in B}$, the second the other half. 
By construction, the two subsets of variables corresponding to the two components are distributed independently and each according to $p^{eq}$.
Without loss of generality, take $i\in A$. Since the two sets $\{\sigma_i\}_{i\in V}$ and $\{\sigma_i\}_{i\in A} \cup \{\sigma'_j\}_{j\in B}$, follow the same distribution, in particular they share the same marginal for the set $\{i\}\cup \partial i$, i.e. $p^{eq}(\sigma_i, \boldsymbol{\sigma}_{\partial i}=\boldsymbol{\sigma'}_{\partial i})=p^{aug}(\sigma_i, \boldsymbol{\sigma'}_{\partial i})=p^{stat}(\boldsymbol{\sigma'}_{\partial i})p^{aug}(\sigma_i| \boldsymbol{\sigma'}_{\partial i})$. 
Marginalizing over the neighbors, one sees that $p^{aug}(\sigma_i)=p^{stat}(\sigma_i)=p^{eq}(\sigma_i)$, thereby proving the first claim.
Moreover, $p^{aug}(\sigma_i| \boldsymbol{\sigma'}_{\partial i})=\tilde{w}(\sigma_{i}^{t+1}=\sigma_i|\boldsymbol{\sigma}_{\partial i}^{t}=\boldsymbol{\sigma'}_{\partial i})$, the transition \eqref{eq:w_glauber}.
By marginalizing over all neighbors but $j$, one obtains that $i$ and $j$ at two subsequent steps of the dynamics follow the equilibrium distribution, proving the second claim.
As acyclic graphs are bipartite, these results hold for any acyclic graph, including the infinite size limits of Erdos-Renyi and Random Regular graphs considered in the article.

Note that the bipartiteness of $G$ is not a serious restriction. Indeed, given an arbitrary graph $G$, possibly non bipartite, one can design a parallel dynamics converging to $p^{eq}$ by considering an associated bipartite graph $G'$ which is constructed from $G$ as follows: for every edge $(i,j)$, add a new spin $\sigma_{ij}$ and replace $(i,j)$ by a couple of edges $(i,(ij)),((ij),j)$ connected to $ij$ with couplings $J_{i,ij}=+\infty, J_{ij,j}=J_{ij}$ (or alternatively, $J_{i,ij}=\tanh^{-1}\left[\sqrt{\tanh(|J_{ij}|)}\right], J_{ij,j}=J_{i,ij}\sign(J_{ij})$). 
Marginalizing over the extra spins $\{\sigma_{ij}\}$ one recovers the original $p^{eq}$ and the new graph $G'$ is clearly bipartite.

\subsection{Self-coupling}
A way of obtaining the equilibrium distribution of a given Ising Hamiltonian that is alternative to the $p_0\to\infty$ limit of \eqref{eq:glauber2} is given by self-couplings.
One can enrich the dynamics by adding a coupling $J_{ii}$ between a spin and itself at the successive epoch. The transition becomes
\begin{equation}
	\tilde{w}(\sigma_{i}^{t+1}|\boldsymbol{\sigma}_{\partial i}^{t},\sigma_i^t)=\frac{e^{\beta\sigma_{i}^{t+1}\left(\sum_{j\in\partial i}J_{ij}\sigma_{j}^{t}+J_{ii}\sigma_i^t+h_{i}\right)}}{2\cosh\left[\beta\left(\sum_{j\in\partial i}J_{ij}\sigma_{j}^{t}+J_{ii}\sigma_i^t+h_{i}\right)\right]}
\end{equation}
and the stationary distribution
\begin{equation}
	p^{stat}(\boldsymbol{\sigma})\propto\exp\left\{\sum_i \left[ \log\cosh\beta\left(\sum_{j\in\partial i} J_{ij}\sigma_j +J_{ii}\sigma_i + h_i\right)+\beta h_i\sigma_i\right]\right\}.  
\end{equation}

In the limit $J_{ii}\gg 1$, one gets
\begin{align}
	\log\cosh\beta\left(\sum_{j\in\partial i} J_{ij}\sigma_j +J_{ii}\sigma_i + h_i\right) = \sum_{j\in\partial i} J_{ij}\sigma_i\sigma_j +J_{ii} + h_i \sigma_i + \mathcal{O}(e^{-J_{ii}})
\end{align} 
and the stationary distribution becomes
\begin{equation}
	p^{stat}(\boldsymbol{\sigma})\propto\exp\left\{2\beta\sum_i \left[ \frac12\sum_{j\in\partial i} J_{ij}\sigma_i\sigma_j + h_i \sigma_i\right]\right\}. 
\end{equation}

By comparison with \eqref{eq:p_ising_eq}, we see that the resulting distribution is that of an Ising model at equilibrium at double the inverse temperature.

\section{Details of the BP equations}

Equation (\ref{eq:exponentialmps}), with matrix indices and the special
cases $t=0,T$ made explicit, reads
\begin{align}
	\left[B_{i\to j}^{0}(x_{i}^{1},x_{i}^{0},x_{j}^{0})\right]_{\{a_{k}^{1}\}_{k\in\partial i\setminus j}} & =\sum_{\{x_{k}^{0}\}_{k\in\partial i\setminus j}}f_{i}^{1}(x_{i}^{1}|\boldsymbol{x}_{\partial i}^{0},x_{i}^{0})\prod_{k\in\partial i\setminus j}\left[A_{k\to i}^{0}(x_{k}^{0},x_{i}^{0})\right]_{a_{k}^{1}}\label{eq:B-1}\\
	\left[B_{i\to j}^{t}(x_{i}^{t+1},x_{i}^{t},x_{j}^{t})\right]_{\{a_{t}^{k},a_{t+1}^{k}\}_{k\in\partial i\setminus j}} & =\sum_{\{x_{k}^{t}\}_{k\in\partial i\setminus j}}f_{i}^{t+1}(x_{i}^{t+1},\boldsymbol{x}_{\partial i}^{t},x_{i}^{t})\prod_{k\in\partial i\setminus j}\left[A_{k\to i}^{t}(x_{k}^{t},x_{i}^{t})\right]_{a_{k}^{t},a_{k}^{t+1}}\quad\forall t\in\{1,\ldots,T-1\}\\
	\left[B_{i\to j}^{T}(x_{i}^{T},x_{j}^{T})\right]_{\{a_{k}^{T}\}_{k\in\partial i\setminus j}} & =\sum_{\{x_{k}^{T}\}_{k\in\partial i\setminus j}}\prod_{k\in\partial i\setminus j}\left[A_{k\to i}^{T}(x_{k}^{T},x_{i}^{T})\right]_{a_{k}^{T}}
\end{align}

\section{How to perform SVD on a tensor}

SVD is only defined for matrices, i.e. arrays with two indices. Whenever
one wishes to apply it to tensors (intended not in the differential-geometric
sense, but as arrays of dimension higher than two), indices must be
split into two subsets and treated as ``macro-indices'' of a new
matrix \cite{oseledets2011tensor}. In computer science lingo, one \emph{reshapes} the high-dimensional
array into a two-dimensional one. For instance, the SVD in (\ref{eq:svd-1})
in full detail reads 
\begin{equation}
	\left[B_{i\to j}^{t}(x_{i}^{t+1},x_{i}^{t},x_{j}^{t})\right]_{\underline{a}^{t},\underline{a}^{t+1}}\stackrel{{\rm SVD}}{=}\sum_{k=1}^{K}\left[C_{i\to j}^{t}(x_{i}^{t},x_{j}^{t})\right]_{\underline{a}^{t},k}\Lambda^t_{kk}\left[V^t(x_{i}^{t+1})\right]^{\dagger}_{k,\underline{a}^{t+1}}
\end{equation}
where $(x_{i}^{t},x_{j}^{t},\underline{a}^{t})$ are treated as a
macro-index for the rows of $B$ and $(x_{i}^{t+1},\underline{a}^{t+1})$
the macro-index for the columns. The range of values for $k$ is determined
by the minimum between the number of rows and columns of $B$: 
\begin{equation}
	K=\min\left\{ q^{2}M^{|\partial i|-1},qM^{|\partial i|-1}\right\} 
\end{equation}
where $q$ is the size of the domain of each $x_i^t$ and $M$ is the bond dimension of the incoming messages, for simplicity
supposed equal for all neighbors and times. Analogously, \eqref{eq:svt-1}
in detail reads
\begin{equation}
	\left[C_{i\to j}^{t}(x_{i}^{t},x_{j}^{t})\right]_{\underline{a}^{t},\underline{a}^{t+1}}\stackrel{{\rm SVD,trunc}}{=}\sum_{k=1}^{M}U^t_{\underline{a}^{t},k}\Lambda^t_{kk}\left[A_{i\to j}^{t}(x_{i}^{t},x_{j}^{t})\right]_{k,\underline{a}^{t+1}}.
\end{equation}
Finally, the orthonormality property (\ref{eq:left_orthogonal}) with
explicit indices reads: 
\begin{equation}
	\sum_{x_{i}^{t},x_{j}^{t},\underline{a}^{t}}\left[C_{i\to j}^{t}(x_{i}^{t},x_{j}^{t})\right]_{\underline{a}^{t},k}\left[C_{i\to j}^{t}(x_{i}^{t},x_{j}^{t})\right]_{\underline{a}^{t},k'}=\delta(k,k')
\end{equation}

\section{Evaluation of observables}

Given a joint distribution in matrix product form
\begin{equation}
	p(x^{0},x^{1},\ldots,x^{T})=\frac{1}{Z}\sum_{a^{1},a^{2},\ldots,a^{T}}\left[A^{0}(x^{0})\right]_{a^{1}}\left[A^{1}(x^{1})\right]_{a^{1},a^{2}}\cdots\left[A^{T-1}(x^{T-1})\right]_{a^{T-1},a^{T}}\left[A^{T}(x^{T})\right]_{a^{T}}
\end{equation}
one can efficiently compute: normalization, marginals, autocorrelations. 

\paragraph{Normalization and marginals}

Marginalizing at time $t$ gives
\begin{align}
	p^{t}(x^{t})= & \sum_{\{x^{s}\}_{s\neq t}}p(x^{0},x^{1},\ldots,x^{T})\\
	=\frac{1}{Z} & \sum_{a^{t},a^{t+1}}\left[L^{t-1}\right]_{a^{t}}\left[A^{t}(x^{t})\right]_{a^{t},a^{t+1}}\left[R^{t+1}\right]_{a^{t+1}}
\end{align}
where we defined partial normalizations from the left and from the
right
\begin{equation}
	\begin{cases}
		\left[L^{t}\right]_{a^{t+1}}\coloneqq & \sum\limits _{\substack{a^{1},\ldots,a^{t}}
		}\prod\limits _{s=0}^{t}\sum\limits _{x^{s}}\left[A^{s}(x^{s})\right]_{a^{s},a^{s+1}}\\
		\left[R^{t}\right]_{a^{t}}\coloneqq & \sum\limits _{\substack{a^{t+1},\ldots,a^{T}}
		}\prod\limits _{s=t}^{T}\sum\limits _{x^{s}}\left[A^{s}(x^{s})\right]_{a^{s},a^{s+1}}
	\end{cases}
\end{equation}
with initial conditions 
\begin{equation}
	\begin{cases}
		\left[L^{0}\right]_{a^{1}}\coloneqq & \sum\limits _{x^{0}}\left[A^{0}(x^{0})\right]_{a^{1}}\\
		\left[R^{T}\right]_{a^{T}}\coloneqq & \sum\limits _{x^{T}}\left[A^{T}(x^{T})\right]_{a^{T}}
	\end{cases}.
\end{equation}

The normalization is given by

\begin{equation}
	Z=\sum_{a^{t}}\left[L^{t}\right]_{a^{t+1}}\left[R^{t+1}\right]_{a^{t+1}}\quad\forall t\in0,1,\ldots,T-1.
\end{equation}

\paragraph{Autocorrelations}

Further define ``middle'' partial normalizations from $t$ to $s$
($t<s$ without loss of generality)

\begin{align}
	\left[M^{t,s}\right]_{a^{t+1},a^{u}} & =\sum_{a^{t+2},\ldots,a^{u-1}}\prod_{u=t+1}^{s-1}\sum_{x_{i}^{u},x_{f}^{u}}\left[A^{u}(x^{u})\right]_{a^{u},a^{u+1}}\\
	& =\sum_{a^{s-1}}\left[M^{t,s-1}\right]_{a^{t+1},a^{s-1}}\left(\sum_{x^{u-1}}\left[A^{s-1}(x^{s-1})\right]_{a^{s-1},a^{s}}\right)
\end{align}
with initial condition 
\begin{equation}
	\left[M^{t,t+1}\right]_{a,b}=\delta(a,b)\quad\forall t\in\{0,1,\ldots,T-1\}.
\end{equation}
Now 
\begin{align}
	p^{t,s}(x^{t},x^{s})= & \sum_{\{x^{u}\}_{u\neq t,s}}p(x^{0},x^{1},\ldots,x^{T})\\
	= & \frac{1}{Z}\sum_{\substack{a^{t},a^{t+1}\\
			a^{s},a^{s+1}
		}
	}\left[L^{t-1}\right]_{a^{t}}\left[A^{t}(x^{t})\right]_{a^{t},a^{t+1}}\left[M^{t,s}\right]_{a^{t+1},a^{s}}\left[A^{s}(x^{s})\right]_{a^{s},a^{s+1}}\left[R^{s+1}\right]_{a^{s+1}}.
\end{align}

\section{Bethe Free Energy}

The Bethe Free Energy for a graphical model with pair-wise interactions
is given by
\begin{equation}
	F_{Bethe}=-\sum_{i}\log z_{i}+\frac{1}{2}\sum_{i}\sum_{j\in\partial i}\log z_{ij}
\end{equation}
where

\begin{align}
	z_{i} & =\sum_{\ox_{i},\boldsymbol{\ox}_{\partial i}}\prod_{t=0}^{T-1}f_{i}^{t+1}(x_{i}^{t+1}|\boldsymbol{x}_{\partial i}^{t},x_{i}^{t})\prod_{k\in\partial i}m_{k\to i}(\ox_{k},\ox_{i})\\
	z_{ij} & =\sum_{\ox_{i},\ox_{j}}m_{i\to j}(\overline{x}_{i},\overline{x}_{j})m_{j\to i}(\overline{x}_{j},\overline{x}_{i}).
\end{align}

It is useful to define

\begin{equation}
	z_{i\to j}=\sum_{\ox_{i},\ox_{j}}\sum_{\boldsymbol{\ox}_{\partial i\setminus j}}\prod_{t=0}^{T-1}f_{i}^{t+1}(x_{i}^{t+1}|\boldsymbol{x}_{\partial i}^{t},x_{i}^{t})\prod_{k\in\partial i\setminus j}m_{k\to i}(\overline{x}_{k},\overline{x}_{i})=\frac{z_{i}}{z_{ij}}.
\end{equation}

Finally,

\begin{align}
	F_{Bethe} & =\sum_{i}\left[\left(\frac{d_{i}}{2}-1\right)\log z_{i}-\frac{1}{2}\sum_{j\in\partial i}\log z_{i\to j}\right]
\end{align}
The Bethe free energy can be obtained using only $\left\{ z_{i}\right\} ,\left\{ z_{i\to j}\right\} ,$which
are already computed when normalizing messages during the BP iterations.

\section{Efficient BP computations\label{sec:Efficient-BP-computations}}

We give here details of the efficient procedure for the computation
of BP messages mentioned in the main text. Re-writing the BP equation
(omitting for clarity the $\phi$ terms) in terms of the auxiliary
variables $\{y_{A}^{t}\}_{A\subseteq\partial i}$ gives
\begin{align}
	m_{i\to j}(\overline{x}_{i},\overline{x}_{j}) & \propto\sum_{\overline{\boldsymbol{x}}_{\partial i\setminus j}}\prod_{t}w(x_{i}^{t+1}|\boldsymbol{x}_{\partial i\setminus j}^{t},x_{i}^{t},x_{j}^{t})\prod_{k\in\partial i\setminus j}m_{k\to i}(\overline{x}_{k},\overline{x}_{i})\\
	& \propto\sum_{\overline{\boldsymbol{x}}_{\partial i\setminus j}}\sum_{\overline{y}_{\partial i\setminus j}}\prod_{t}p(x_{i}^{t+1}|y_{\partial i\setminus j}^{t},x_{i}^{t},x_{j}^{t})p(y_{\partial i\setminus j}^{t}|\boldsymbol{x}_{\partial i\setminus j}^{t},x_{i}^{t})\prod_{k\in\partial i\setminus j}m_{k\to i}(\overline{x}_{k},\overline{x}_{i})\\
	& \propto\sum_{\overline{y}_{\partial i\setminus j}}\prod_{t}p(x_{i}^{t+1}|y_{\partial i\setminus j}^{t},x_{i}^{t},x_{j}^{t})\tilde{m}_{\partial i\setminus j\to i}(\overline{y}_{\partial i\setminus j},\overline{x}_{i})\label{eq:msg_last_step}
\end{align}
where we defined $\tilde{m}_{\partial i\setminus j\to i}(\overline{y}_{\partial i\setminus j},\overline{x}_{i})=\sum_{\boldsymbol{\overline{x}}_{\partial i\setminus j}}\prod_{t}p(y_{\partial i\setminus j}^{t}|\boldsymbol{x}_{\partial i\setminus j}^{t},x_{i}^{t})\prod_{k\in\partial i\setminus j}m_{k\to i}(\overline{x}_{k},\overline{x}_{i})$. 

Now $\tilde{m}_{\partial i\setminus j\to i}$ can be computed as the
aggregation of all messages $\tilde{m}_{k\to i}$with $k<j$ and messages
$\tilde{m}_{k\to i}$with $k>j$:

\begin{equation}
	\tilde{m}_{\partial i\setminus j\to i}(\overline{y}_{\partial i\setminus j},\overline{x}_{i})=\sum_{\overline{y}_{<j}}\sum_{\overline{y}_{>j}}\prod_{t}p(y_{\partial i\setminus j}^{t}|y_{<j}^{t},y_{>j}^{t},x_{i}^{t})\tilde{m}_{<j}(\overline{y}_{<j},\overline{x}_{i})\tilde{m}_{>j}(\overline{y}_{>j},\overline{x}_{i})\label{eq:bp_efficient_msg}
\end{equation}
where we used the short-hand notation $\lessgtr j=\{k\in\partial i\setminus j,k\lessgtr j\}$.
The last equation is naturally cast to matrix product form with 
\begin{equation}
	\left[\tilde{A}_{\partial i\setminus j\to i}^{t}(y_{\partial i\setminus j}^{t},x_{i}^{t})\right]_{(a^{t},b^{t}),(a^{t+1},b^{t+1})}=\sum_{y_{<j}^{t}}\sum_{y_{>j}^{t}}p(y_{\partial i\setminus j}^{t}|y_{<j}^{t},y_{>j}^{t},x_{i}^{t})\left[\tilde{A}_{<j}(y_{<j}^{t},x_{i}^{t})\right]_{a^{t},a^{t+1}}\left[\tilde{A}_{>j}(y_{>j}^{t},x_{i}^{t})\right]_{b^{t},b^{t+1}}
\end{equation}
where subscripts for the matrices match those of the corresponding
messages in (\ref{eq:bp_efficient_msg}). Matrices on the LHS have
size double than those at RHS, therefore we perform the same SVD-based
truncations explained in the main text. This is where the computational
bottleneck lies: suppose that the incoming matrices have size $M\times M$.
Performing a SVD on $\tilde{A}_{\partial i\setminus j\to i}^{t}$,
reshaped as a matrix with $(a^{t},b^{t})$ as row index and $(a^{t+1},b^{t+1},y_{\partial i\setminus j}^{t},x_{i}^{t})$
as column index, costs $\mathcal{O}(nM^{6})$ where $n$ is the number
of values taken by $y_{\partial i\setminus j}^{t}$ and depends on
the model. As long as $n$ depends at most polynomially on the degree
$z=|\partial i|$, the exponential dependence is avoided. 

Messages $\tilde{m}$ can be computed recursively after having noticed
that they satisfy analogous properties to (\ref{eq:pAuB}): 
\begin{equation}
	\tilde{m}_{A\cup B}(\overline{y}_{A\cup B},\overline{x}_{i})=\sum_{\overline{y}_{A},\overline{y}_{B}}\prod_{t}p(y_{A\cup B}^{t}|y_{A}^{t},y_{B}^{t},x_{i}^{t})\tilde{m}_{A}(\overline{y}_{A},\overline{x}_{i})\tilde{m}_{B}(\overline{y}_{B},\overline{x}_{i})\label{eq:recursion_mtilde}
\end{equation}
starting from $\tilde{m}_{\{k\}\to i}(\overline{y}_{\{k\}},\overline{x}_{i})=\sum_{\overline{x}_{k}}\prod_{t}p(y_{\{k\}}^{t}|x_{k}^{t},x_{i}^{t})m_{k\to i}(\overline{x}_{k},\overline{x}_{i})$
and $\tilde{m}_{\emptyset\to i}(\overline{y}_{\emptyset},\overline{x}_{i})\propto1\:\forall\:(\overline{y}_{\emptyset},\overline{x}_{i}$).
Finally, we use (\ref{eq:msg_last_step}) to compute $m_{i\to j}(\overline{x}_{i},\overline{x}_{j})$
for all $j$: just as in (\ref{eq:exponentialmps}) we get matrices
with dependency on both $x_{i}^{t+1}$ and $x_{i}^{t}$
\begin{equation}
	B_{i\to j}^{t}(x_{i}^{t+1},x_{i}^{t},x_{j}^{t})=\sum_{y_{\partial i\setminus j}^{t}}p(x_{i}^{t+1}|y_{\partial i\setminus j}^{t},x_{i}^{t},x_{j}^{t})\tilde{A}_{\partial i\setminus j\to i}(y_{\partial i\setminus j}^{t},x_{i}^{t})\label{eq:B_efficient}
\end{equation}
which are treated in the same way as explained in the main text for
the generic BP implementation. At this point one can use the already
computed quantities to retrieve the belief at node $i$

\begin{equation}
	b_{i}(\overline{x}_{i})\propto\sum_{\overline{y}_{\partial i}}\prod_{t}p(x_{i}^{t+1}|y_{\partial i}^{t},x_{i}^{t})\tilde{m}_{\partial i\to i}(\overline{y}_{\partial i},\overline{x}_{i})
\end{equation}
with $j$ being any neighbor of $i$.

The strategy just described is summarized in algorithm \ref{alg:Efficient-computation}.
The procedure is manifestly linear in the degree, for an overall cost
of $\mathcal{O}(znM^{6})$ for the update of all messages outgoing
from a node. In cases where there exists no convenient choice for
the auxiliary variables $y$, the scheme could still be implemented
with $y_{A}^{t}=\otimes_{a\in A}\{x_{A}^{t}\}$ and $n\sim q^{z}$:
unsurprisingly, one recovers the exponential cost with respect to
the degree.

\begin{algorithm}[H]
	\begin{itemize}
		\item for $j\in\partial i$
		\begin{itemize}
			\item $\tilde{m}_{\{k\}\to i}(\overline{y}_{\{k\}},\overline{x}_{i})\leftarrow\sum_{\overline{x}_{k}}\prod_{t}p(y_{\{k\}}^{t}|x_{k}^{t},x_{i}^{t})m_{k\to i}(\overline{x}_{k},\overline{x}_{i})$
		\end{itemize}
		\item $\tilde{m}_{\emptyset\to i}(\overline{y}_{\emptyset},\overline{x}_{i})\leftarrow1$
		\item for $j\in\partial i$
		\begin{itemize}
			\item $\tilde{m}_{<j}(\overline{y}_{<j},\overline{x}_{i})\leftarrow\sum_{\overline{y}_{<j-1}}\sum_{\overline{y}_{\{j-1\}}}\prod_{t}p(y_{<j}^{t}|y_{<j-1}^{t},y_{\{j-1\}}^{t},x_{i}^{t})\tilde{m}_{<j-1}(\overline{y}_{<j-1},\overline{x}_{i})\tilde{m}_{\{j-1\}}(\overline{y}_{\{j-1\}},\overline{x}_{i})$
			\item $\tilde{m}_{>j}(\overline{y}_{>j},\overline{x}_{i})\leftarrow\sum_{\overline{y}_{>j+1}}\sum_{\overline{y}_{\{j+1\}}}\prod_{t}p(y_{>j}^{t}|y_{>j+1}^{t},y_{\{j+1\}}^{t},x_{i}^{t})\tilde{m}_{>j+1}(\overline{y}_{>j+1},\overline{x}_{i})\tilde{m}_{\{j+1\}}(\overline{y}_{\{j+1\}},\overline{x}_{i})$
		\end{itemize}
		\item for $j\in\partial i$
		\begin{itemize}
			\item $\tilde{m}_{\partial i\setminus j\to i}(\overline{y}_{\partial i\setminus j},\overline{x}_{i})\leftarrow\sum_{\overline{y}_{<j}}\sum_{\overline{y}_{>j}}\prod_{t}p(y_{\partial i\setminus j}^{t}|y_{<j}^{t},y_{>j}^{t},x_{i}^{t})\tilde{m}_{<j}(\overline{y}_{<j},\overline{x}_{i})\tilde{m}_{>j}(\overline{y}_{>j},\overline{x}_{i})$
			\item $m_{i\to j}(\overline{x}_{i},\overline{x}_{j})\leftarrow\sum_{\overline{y}_{\partial i\setminus j}}\prod_{t}p(x_{i}^{t+1}|y_{\partial i\setminus j}^{t},x_{i}^{t},x_{j}^{t})\tilde{m}_{\partial i\setminus j\to i}(\overline{y}_{\partial i\setminus j},\overline{x}_{i})$%
		\end{itemize}
	\end{itemize}
	\caption{\label{alg:Efficient-computation}Efficient computation of outgoing
		messages and belief for a generic node $i$.}
\end{algorithm}

\begin{figure*}[htb]
	\includegraphics[width=1\textwidth]{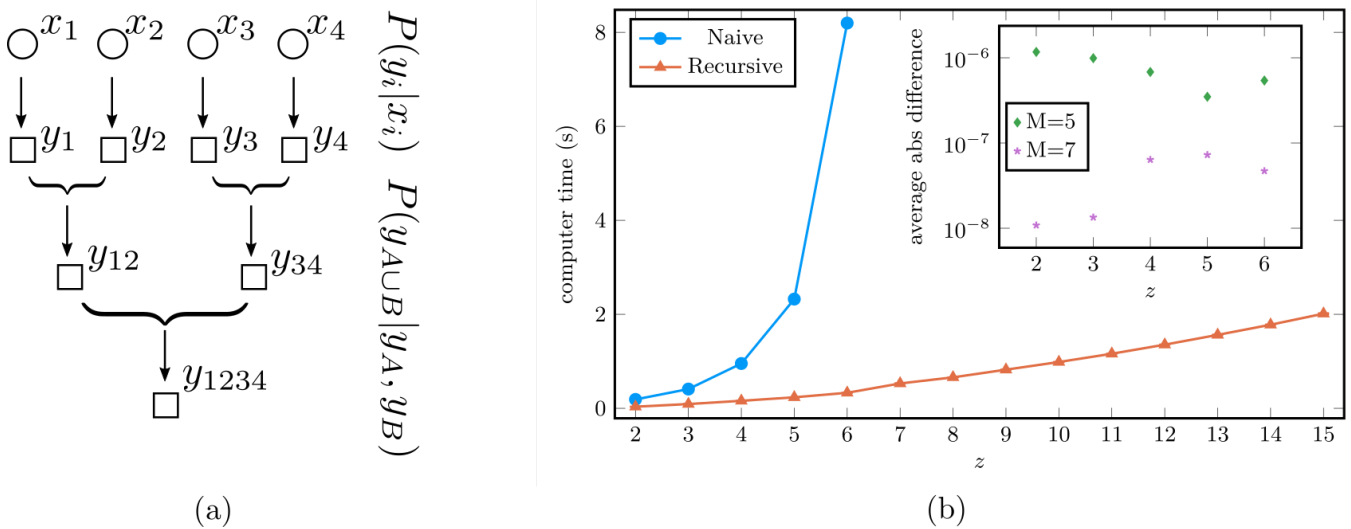}
	\caption{\label{fig:recursive} (a) Sketch of the recursive procedure described in this section. (b) Computer time to run 10 iterations of MPBP with the naive vs recursive update for a SIS model on a star graph of degree $z$, $\lambda = 0.2, \rho = 0.1, \gamma=0.05$, no reweighting, bond dimension $5$, average over $20$ random instances. Error bars are smaller than the points. Inset: absolute difference between values of the marginals for the two methods, averaged over epochs, sites and instances, for two values of bond dimension. Such very small discrepancies are due to the fact that the recursive update, unlike the naive one, performs truncations at each intermediate step.}
\end{figure*}

Figure \ref{fig:recursive} sketches the recursive procedure described above and shows the computation time necessary to run $10$ iterations of MPBP for a SIS model on a star graph (one central node connected to $z$ others) of varying size.
The naive update scheme shows exponential growth in computational time, in contrast with the linear behavior of the recursive strategy.  

For the SIS model (SIRS behaves analogously) we pick $y_{A}^{t}$ to be the event that at least
one of $k\in A$ infects $i$:
\begin{align}
	p\left(y_{k}^{t}|x_{k}^{t},x_{i}^{t}\right) & =\begin{cases}
		\lambda_{ki}\delta(y_{j}^{t},I)+(1-\lambda_{ki})\delta(y_{k}^{t},S) & \text{if }x_{k}^{t}=S\\
		\delta(y_{k}^{t},S) & \text{otherwise }
	\end{cases}\\
	p\left(y_{A\cup B}^{t}|y_{A}^{t},y_{B}^{t},x_{i}^{t}\right) & =\delta(y_{A\cup B}^{t},I)\mathds{1}\left[y_{A}^{t}=I\vee y_{B}^{t}=I\right]+\delta(y_{A\cup B}^{t},S)\mathds{1}\left[y_{A}^{t}=S\wedge y_{B}^{t}=S\right]
\end{align}
where $\mathds{1}\left[\mathcal{P}\right]$ is the indicator function which evaluates to $1$ when predicate $\mathcal{P}$ is true, to $0$ otherwise.

In this case, all $y$ variables are binary, yielding a computational
cost $\mathcal{O}(zM^{6})$ for the update of $z$ messages.

In the case of parallel Glauber dynamics the most general setting where these simplifications
apply is couplings with constant absolute value $\left|J_{ij}\right|\equiv J$
and arbitrary external fields, often referred to as the $\pm J$ Ising
model. The case with $J_{ij}\equiv J$, $h=0$ studied in \citep{barthel2018matrix}
is automatically covered. The transition probability (\ref{eq:w_glauber})
takes the form
\begin{align}
	e^{\beta\sigma_{i}^{t+1}\left[J\left(\sum_{j\in\partial i}\sign(J_{ij})\sigma_{j}^{t}\right)+h_{i}\right]} & \propto e^{\beta\sigma_{i}^{t+1}\left[J\left(y_{\partial i\setminus j}^{t}+\sign(J_{ij})\sigma_{j}^{t}\right)+h_{i}\right]}
\end{align}
with $y_{A}^{t}=\sum_{k\in A}\sign(J_{ik})\sigma_{k}^{t}$. It is
easy to see that $p\left(y_{\{k\}}^{t}|\sigma_{k}^{t},\sigma_{i}^{t}\right)=\delta\left(y_{\{k\}}^{t},\sign(J_{ik})\sigma_{k}^{t}\right)$
and $p\left(y_{A\cup B}^{t}|y_{A}^{t},y_{B}^{t},\sigma_{i}^{t}\right)=\delta\left(y_{A\cup B}^{t},y_{A}^{t}+y_{B}^{t}\right)$.
In this case, $y_{A}^{t}$ can take value $-|A|,-|A|+2,\ldots,|A|-2,|A|$,
for a total $2|A|+1$ values. The maximum is achieved for $A=\partial i\setminus j$,
yielding a computational cost $\mathcal{O}(z^{2}M^{6})$ for the update
of $z$ messages.

\section{Population dynamics}
For systems with homogeneous properties (e.g. Ising model on a regular graph with homogeneous coupling constant $J_{ij}\equiv J$ and external field $h_i\equiv h$), efficient computations in the thermodynamic limit $N\to\infty$ are possible (see e.g. fig. \ref{fig:glauber}(a)).
Messages living on each edge of the graph asymptotically become all equal, therefore it is enough to store a single message. This is a standard approach within the cavity method \cite{mezard2009information} and has been used also in \cite{barthel2018matrix}.
Whenever the node degree or other parameters of the system are distributed according to some disorder, such a simple approach is not viable. The standard strategy in these cases is to work with a finite collection of BP messages playing the role of a discretized approximation to the true distribution of messages within the disorder ensemble. The approach is called population dynamics \cite{mezard2009information} and has been used in this paper to produce the data in figure \ref{fig:glauber}(b) where the node degree is randomly distributed.

A population of $P$ messages in matrix-product form \eqref{eq:mps} is initialized at random.
Then, the following is iterated a sufficiently large number $N_{it}$ of times as follows.
At each iteration, a degree $z$ is sampled from the degree distribution, then $z$ messages are picked at random from the population. At this point one can imagine a node with $z$ neighbors and the picked messages incoming through the $z$ edges. The outgoing messages are computed according to the BP equation \eqref{eq:bp}, with SVD truncations to some fixed bond dimension.
With little further computational effort, the belief (marginal probability distribution \eqref{eq:belief}) and possibly other observables are also calculated and stored.
The newly computed messages are then inserted into the population replacing the ones used as incoming.
After $N_{it}$ such iterations, the output of the algorithm is the statistics over the stored observables.
Care must be taken in selecting only the samples collected after a stationary state has been reached, i.e. when the population had converged to a good representation of the target probability distribution.

The whole procedure can be run multiple times with increasing bond dimension to verify whether a better approximation can be achieved with larger matrices.
The bond dimension is in principle allowed to vary also within the iterations.

A pseudo-code implementation for Glauber dynamics on an infinite Erdos-Renyi graph is provided in algorithm \ref{alg:popdyn}.

\begin{algorithm}[H]
	\caption{Population dynamics for Glauber dynamics on infinite Erdos-Renyi graph, pseudo-code}\label{alg:popdyn}
	\begin{algorithmic}
		\State $\boldsymbol{m}$: array of $P$ randomly-initialized messages
		\State $\boldsymbol{\tilde{m}}$: auxiliary array of messages for intermediate calculations		
		\State $p(z)$: residual degree distribution (Poisson)
		\State $N_{it}$: number of iterations
		\State $M$: max bond dimension for SVD truncations
		
%		\State $n=0$
		\For{$it \in \{1,2,\ldots,N_{it}\}$}
			\State sample $z\sim p(z)$
			\State sample $i_1,i_2,\ldots,i_z$ from $\{1,2,\ldots,P\}$
			\For{$j\in \{i_1,i_2,\ldots,i_z\}$}
			\State $\tilde{m}_j \gets f_{BP}(\boldsymbol{m}_{\{i_1,i_2,\ldots,i_z\}\setminus j})$, truncations to size $M$ \Comment{$f_{BP}$ is \eqref{eq:bp}}
			\EndFor
			\For{$j\in \{i_1,i_2,\ldots,i_z\}$}
			\State $m_j \gets \tilde{m}_j $
			\EndFor
%			\If{$\boldsymbol{m}$ has converged}
			\State store $b=f_{\rm belief}(\boldsymbol{m}_{\{i_1,i_2,\ldots,i_z\}})$ \Comment{$f_{\rm belief}$ is \eqref{eq:belief}}
%			\State $n \gets n + 1$
%			\EndIf
		\EndFor
		\State \textbf{Output}: average over the stored beliefs, used to estimate average magnetization and autocovariance
	\end{algorithmic}
\end{algorithm}

\section{Discretized mean-field methods}

We report the expressions for the discretized version of Dynamic Message Passing (DMP), Individual-Based
Mean Field (IBMF) and Cavity Master Equation (CME) which were used
to produce the data in fig. \ref{fig:sis_free}. They consist in a
discrete time evolution for the expectation of single-variable marginals
and cavity marginals (DMP and CME). In the limit of infinitesimal time-step,
they reduce to their continuous counterparts. Define $I_{i}^{t}$
as the probability of individual $i$ being in state $I$ at time
$t$, $I_{i\to j}^{t}$ ($(ij)\in E$) the probability of individual
$i$ being in state $I$ at time $t$ and having been infected by
someone other than $j$. We parametrize transmission and recovery
probabilities as a rate $\lambda,\rho$ times the time-step $\Delta t$
so that in the continuous-time limit, the equations in their original
version are recovered in terms of rates.

For IBMF we have

\begin{equation}
	I_{i}^{t+\Delta t}=(1-\rho_{i}\Delta t)I_{i}^{t}+\left(1-\prod_{j\in\partial i}(1-\lambda_{ji}\Delta tI_{j}^{t})\right)(1-I_{i}^{t})
\end{equation}
for DMP
\begin{align}
	I_{i}^{t+\Delta t} & =(1-\rho_{i}\Delta t)I_{i}^{t}+\left(1-\prod_{j\in\partial i}(1-\lambda_{ji}\Delta tI_{j\to i}^{t})\right)(1-I_{i}^{t})\\
	I_{i\to j}^{t+\Delta t} & =(1-\rho_{i}\Delta t)I_{i\to j}^{t}+\left(1-\prod_{k\in\partial i\setminus j}(1-\lambda_{ki}\Delta tI_{k\to i}^{t})\right)(1-I_{i}^{t})
\end{align}
and for CME
\begin{align}
	I_{i}^{t+\Delta t} & =(1-\rho_{i}\Delta t)I_{i}^{t}+\left(1-\prod_{j\in\partial i}(1-\lambda_{ji}\Delta tI_{j\to i}^{t})\right)(1-I_{i}^{t})\\
	I_{i\to j}^{t+\Delta t} & =(1-\rho_{i}\Delta t)I_{i\to j}^{t}+\left(1-\prod_{k\in\partial i\setminus j}(1-\lambda_{ki}\Delta tI_{k\to i}^{t})\right)(1-I_{i\to j}^{t})
\end{align}

\section{Exact mappings}

We show examples of models which can be represented exactly by a MPS.

\paragraph{Models with mass on a finite support}

Any arbitrary distribution $p(\overline{x})=p(x^{0},x^{1},\ldots,x^{T})$
can in principle be represented via a MPS, albeit with bond dimension
exponentially large in $T$: to see this, re-write $p$ trivially
as a superposition of delta distributions

\begin{equation}
	p(\overline{x})=\sum_{\overline{y}}p(\overline{y})\prod_{t=0}^{T}\delta\left(x^{t},y^{t}\right)\label{eq:mps_deltas}
\end{equation}
where the product over $t$ is interpreted as a product of $1\times1$
matrices. Since the linear combination of two MPSs is itself a MPS \cite{oseledets2011tensor}:

\begin{equation}
	a\prod_{t}A^{t}(x^{t})+b\prod_{t}B^{t}(x^{t})=\prod_{t}C^{t}(x^{t})\label{eq:sum_of_mps}
\end{equation}
with
\begin{equation}
	C^{0}(x^{0})=\left[\begin{array}{cc}
		aA^{0}(x^{0}) & bB^{0}(x^{0})\end{array}\right],\quad C^{t}(x^{t})=\left[\begin{array}{cc}
		A^{t}(x^{t}) & 0\\
		0 & B^{t}(x^{t})
	\end{array}\right],\quad C^{T}(x^{T})=\left[\begin{array}{c}
		A^{T}(x^{T})\\
		B^{T}(x^{T})
	\end{array}\right]
\end{equation}
then $p$ can be expressed by a MPS with bond dimension $q^{T}$, $q$
being the number of values taken by each $x^{t}$. Now, if the distribution
under consideration puts non-zero probability only over a small set
$\mathcal{T}$ of trajectories, the number of components in the superposition,
and hence the final bond dimension, is $|\mathcal{T}|.$ 

Any non-recurrent and Markovian model with $q$ states such as SIR
(Susceptible Infectious Recovered, $q=3$), SEIR (Susceptible Exposed
Infectious Recovered, $q=4$), etc., allows only a sub-exponential
fraction of the $q^{T}$ potential trajectories. Take as an example
the SIR model: each message $m_{i\to j}$ can be parametrized by the
infection and recovery times for individuals $i$ and $j$, for a
total $\mathcal{O}(T^{4})$ possible trajectories. The same reasoning
goes for a generic non-recurrent Markovian model with $q$ states,
yielding bond dimension $\mathcal{O}(T^{q^{2}})$.

\paragraph{Chain models}

Consider $T+1$ variables each taking one in $q$ values whose distribution
is factorized over an open chain
\begin{equation}
	p(x^{0},x^{1},\ldots,x^{T})\propto\prod_{i=0}^{T-1}\psi^{t}(x^{t},x^{t+1}).
\end{equation}

We show that there exists an equivalent formulation in MPS form, with
matrices of size $q\times q$. Introduce additional variables $\{a^{t}\}_{t=1:T}$with
$a^{t}=x^{t}$ to get

\begin{align}
	p(x^{1},x^{2},\ldots,x^{T}) & \propto\sum_{a^{1},a^{2},\ldots,a^{T}}\delta(x^{0},a^{1})\prod_{t=0}^{T-2}\left\{ \psi^{t}(a^{t+1},x^{t+1})\delta(x^{t+1},a^{t+2})\right\} \psi^{T-1}(a^{T},x^{T})\\
	& \propto\sum_{a^{1},a^{2},\ldots,a^{T}}\left[A^{0}(x^{0})\right]_{a^{1}}\prod_{t=1}^{T-1}\left[A^{t}(x^{t})\right]_{a^{t},a^{t+1}}\left[A^{T}(x^{T})\right]_{a^{T}}\\
	& \propto\prod_{t=0}^{T}A^{t}(x^{t})
\end{align}
with

\begin{equation}
	\begin{cases}
		\left[A^{0}(x^{0})\right]_{a^{1}} & =\delta(x^{0},a^{1})\\
		\left[A^{t}(x^{t})\right]_{a^{t},a^{t+1}} & =\psi^{t-1}(a^{t},x^{t})\delta(x^{t},a^{t+1})\\
		\left[A^{N}(x^{T})\right]_{a^{N}} & =\psi^{T-1}(a^{T},x^{T})
	\end{cases}\quad\forall t\in1,2,\ldots,T-1
\end{equation}
where each $a$ ranges over $q$ values. We note the following implication:
messages in the $1$-step DMP method \citet{delferraro2015dynamic},
which are parametrized as chain models, can be represented with matrices
of size $q^{2}\times q^{2}$.

\paragraph{One-particle trajectories in the SI model}

We show that the probability of any trajectory of an individual in
the SI model can be represented by a MPS with matrices of size $2\times2$.
It suffices to show that such probability factorizes over a chain.
In the following we will sometimes use the convention $S=0,I=1$.
The rule that once an individual $i$ is infected at time $t$ it
can never recover is then encoded compactly as $\prod_{t=1}^{T}\mathds{1}\left[x_{i}^{t+1}\ge x_{i}^{t}\right]$. 

For a generic time $t$ consider the conditional probability $p(x^{t+1}|x^{0},x^{1},\ldots,x^{t})$.
If $x^{t}=I$ then $p(x^{t+1}|x^{0},x^{1},\ldots,x^{t})=\delta(x^{t+1},I)$.
If $x^{t}=S$ then it must also be that $x^{0}=x^{1}=\ldots=x^{t-1}=S$.
We conclude that the state at time $t+1$ depends on the previous
states only through the state at time $t$: $p(x^{t+1}|x^{0},x^{1},\ldots,x^{t})=p(x^{t+1}|x^{t})$.
Hence,

\begin{equation}
	p(x^{0},x^{1},\ldots,x^{N})=\prod_{t=0}^{T-1}p(x^{t+1}|x^{0},x^{1},\ldots,x^{t})=\prod_{t=0}^{T-1}p(x^{t+1}|x^{t})
\end{equation}
The same thesis can be proven via a different argument: for ``non-recurrent''
models like SI, information about the trajectory can be encoded into
a single parameter: the infection time. Infection at some time $t_{i}\in\{0,1,\ldots,T,\infty\}$
(we use the convention that no infection corresponds to $t=\infty$)
corresponds to $x^{0}=\ldots=x^{t-1}=S,x^{t}=\ldots,x^{T}=I$. It
is sometimes convenient to switch between these two equivalent representations.

We propose a chain-factorized ansatz and show that it fully specifies
the probability of a trajectory

\begin{equation}
	p(x^{0},x^{1},\ldots,x^{T})=\left[\prod_{t=0}^{T-1}\mathds{1}\left[x^{t}\le x^{t+1}\right]q^{t}(x^{t})\right]q^{T}(x^{T}).
\end{equation}

The probability of any of the allowed trajectories is

\begin{equation}
	p(t_{i}=t)=p(x^{0}=\ldots=x^{t-1}=S,x^{t}=\ldots,x^{T}=I)=\prod_{t=0}^{t-1}q^{t}(S)\prod_{t=t}^{T}q^{t}(I).
\end{equation}

The ratio of probabilities of infection at times $t+1$ and $t$ gives

\begin{equation}
	\frac{p(t_{i}=t+1)}{p(t_{i}=t)}=\frac{q^{t}(S)}{q^{t}(I)}.
\end{equation}

Parametrizing as $q^{t}(S)\propto1,q^{t}(I)\propto e^{-h^{t}}$, we
get

\begin{equation}
	h^{t}=\log\frac{q^{t}(S)}{q^{t}(I)}=\log\frac{p(t_{i}=t+1)}{p(t_{i}=t)}.
\end{equation}

In full detail, the resulting MPS is

\begin{equation}
	\begin{cases}
		\left[A^{0}(x^{0})\right]_{a^{1}} & =\delta(x^{0},a^{1})\\
		\left[A^{t}(x^{t})\right]_{a^{t},a^{t+1}} & =\mathds{1}\text{\ensuremath{\left[a^{t}\le x^{t}\right]}}q^{t-1}(a^{t})\delta(x^{t},a^{t+1})\quad\forall t\in1,\ldots,T-1\\
		\left[A^{T}(x^{T})\right]_{a^{T}} & =\mathds{1}\text{\ensuremath{\left[a^{T}\le x^{T}\right]}}q^{T-1}(a^{T})q^{T}(x^{T})
	\end{cases}
\end{equation}

\paragraph{Pair trajectories in the SI model}

We show that any BP message in the SI model can be represented exactly
by a MPS with matrices of size $6\times6$. Consider the BP equations
for the SI model parametrized with infection times (see \cite{altarelli2014bayesian})

\begin{align}
	m_{i\to j}(t_{i},t_{j}) & \propto\sum_{t_{\partial i\setminus j}}\delta\left(t_{i},\min_{k\in\partial i}\left\{ t_{k}\right\} \right)\prod_{k\in\partial i\setminus j}m_{k\to i}(t_{k},t_{i})\\
	& \propto\mathds{1}\left[t_{i}\le t_{j}\right]\prod_{k\in\partial i\setminus j}\sum_{t_{k}}\mathds{1}\left[t_{i}\le t_{k}\right]m_{k\to i}(t_{k},t_{i})-\mathds{1}\left[t_{i}<t_{j}\right]\prod_{k\in\partial i\setminus j}\sum_{t_{k}}\mathds{1}\left[t_{i}<t_{k}\right]m_{k\to i}(t_{k},t_{i})\\
	& \propto\mathds{1}\left[t_{i}\le t_{j}\right]a_{i\to j}(t_{i})-\mathds{1}\left[t_{i}<t_{j}\right]b_{i\to j}(t_{i})\\
	& \propto\mathds{1}\left[t_{i}\le t_{j}\right]c_{i\to j}(t_{i})+\mathds{1}\left[t_{i}=t_{j}\right]b_{i\to j}(t_{i})
\end{align}
where we used $\delta\left(x,\min_{k\in S}\left\{ x_{k}\right\} \right)=\prod_{k\in S}\mathds{1}\left[x\le x_{k}\right]-\prod_{k\in S}\mathds{1}\left[x<x_{k}\right]$
and defined $a_{i\to j}(t_{i})=\prod_{k\in\partial i\setminus j}\sum_{t_{k}}\mathds{1}\left[t_{i}\le t_{k}\right]m_{k\to i}(t_{k},t_{i})$,
$b_{i\to j}(t_{i})=\prod_{k\in\partial i\setminus j}\sum_{t_{k}}\mathds{1}\left[t_{i}<t_{k}\right]m_{k\to i}(t_{k},t_{i})$,
$c_{i\to j}(t_{i})=a_{i\to j}(t_{i})-b_{i\to j}(t_{i})$.

Once normalized, both $c_{i\to j}$ and $b_{i\to j}$ are probability
distributions for a single SI trajectory, hence they can be re-parametrized
(with a slight abuse of notation) as MPSs $c_{i\to j}(x_{i})=\prod_{t}\mathds{1}\left[x_{i}^{t+1}\ge x_{i}^{t}\right]\tilde{c}_{i\to j}^{t}(x_{i}^{t})$,
$b_{i\to j}(x_{i})=\prod_{t}\mathds{1}\left[x_{i}^{t+1}\ge x_{i}^{t}\right]\tilde{b}_{i\to j}^{t}(x_{i}^{t})$.
Introducing the SI rule also for $x_{j},$we get

\begin{equation}
	m_{i\to j}(x_{i},x_{j})\propto\prod_{t}\mathds{1}\left[x_{i}^{t}=x_{j}^{t}\right]\mathds{1}\left[x_{i}^{t+1}\ge x_{i}^{t}\right]\tilde{b}_{i\to j}(x_{i})+\prod_{t}\mathds{1}\left[x_{i}^{t}\le x_{j}^{t}\right]\mathds{1}\left[x_{i}^{t+1}\ge x_{i}^{t}\right]\mathds{1}\left[x_{j}^{t+1}\ge x_{j}^{t}\right]\tilde{c}_{i\to j}^{t}(x_{i}^{t}).
\end{equation}

The first term is a chain-factorized distribution for, say, $x_{i}$
times the constraint $x_{j}^{t}=x_{i}^{t}\forall t$, hence it can
be represented as an MPS with $2\times2$ matrices. The second term
is a chain of $4$-state variables $\{(x_{i}^{t},x_{j}^{t})\}_{t=0:T}$,
hence it can be represented as an MPS with $4\times4$ matrices. In
full detail

\begin{equation}
	\prod_{t}\mathds{1}\left[x_{i}^{t}=x_{j}^{t}\right]\mathds{1}\left[x_{i}^{t+1}\ge x_{i}^{t}\right]\tilde{b}_{i\to j}(x_{i})=\sum_{a_{i}^{1},\ldots,a_{i}^{T}}\prod_{t}\underbrace{\mathds{1}\left[x_{i}^{t}=x_{j}^{t}\right]\delta(x_{i}^{t},a_{i}^{t+1})\mathds{1}\left[a_{i}^{t}\le x_{i}^{t}\right]\tilde{b^{t-1}}_{i\to j}(a_{i}^{t})}_{\left[B^{t}(x_{i}^{t},x_{j}^{t})\right]_{a_{i}^{t},a_{i}^{t+1}}}
\end{equation}

\begin{equation}
	\prod_{t}\mathds{1}\left[x_{i}^{t}\le x_{j}^{t}\right]\mathds{1}\left[x_{i}^{t+1}\ge x_{i}^{t}\right]\mathds{1}\left[x_{j}^{t+1}\ge x_{j}^{t}\right]\tilde{c}_{i\to j}^{t}(x_{i}^{t})=\sum_{\substack{a_{i}^{1},\ldots,a_{i}^{T}\\
			a_{j}^{1},\ldots,a_{j}^{T}
		}
	}\prod_{t}\underbrace{\mathds{1}\left[x_{i}^{t}\le x_{j}^{t}\right]\delta(x_{i}^{t},a_{i}^{t+1})\delta(x_{j}^{t},a_{j}^{t+1})\mathds{1}\left[a_{i}^{t}\le x_{i}^{t}\right]\mathds{1}\left[a_{j}^{t}\le x_{j}^{t}\right]\tilde{c}_{i\to j}^{t-1}(a_{i}^{t})}_{\left[C^{t}(x^{t})\right]_{(a_{i}^{t},a_{j}^{t}),(a_{i}^{t+1},a_{j}^{t+1})}}
\end{equation}

Finally, since the mixture of two MPSs is itself an MPS (\ref{eq:sum_of_mps}),
we get that $m_{i\to j}$ can be written as a MPS with matrices of
size $2+4=6$.

\section{Pair-wise reweightings}

The distribution (\ref{eq:p}) can be made more general by adding
reweighting terms involving neighboring variables $\left\{ \psi_{ij}^{t}(x_{i}^{t},x_{j}^{t})\right\} _{(ij)\in E}$.
Now 
\begin{equation}
	\begin{aligned}p(\overline{\boldsymbol{x}})\propto & \prod_{i=1}^{N}w(x_{i}^{0})\prod_{t=0}^{T-1}\prod_{i=1}^{N}w(x_{i}^{t}|\boldsymbol{x}_{\partial i}^{t-1},x_{i}^{t-1})\phi_{i}^{t}(x_{i}^{t})\prod_{(ij)}\psi_{ij}^{t}(x_{i}^{t},x_{j}^{t})\end{aligned}
	.
\end{equation}

The message ansatz stays the same. The BP equation becomes

\begin{equation}
	m_{i\to j}(\overline{x}_{i},\overline{x}_{j})\propto\sum_{\overline{x}_{\partial i\setminus j}}w(x_{i}^{0})\phi_{i}^{0}(x_{i}^{0})\prod_{t}w(x_{i}^{t+1}|\boldsymbol{x}_{\partial i}^{t},x_{i}^{t})\phi_{i}^{t+1}(x_{i}^{t+1})\prod_{k\in\partial i\setminus j}\psi_{ik}^{t+1}(x_{i}^{t+1},x_{k}^{t+1})\prod_{k\in\partial i\setminus j}m_{k\to i}(\overline{x}_{k},\overline{x}_{i})
\end{equation}
and the $B$ matrices read

\begin{equation}
	\begin{aligned}\left[B_{i\to j}^{0}(x_{i}^{1},x_{i}^{0},x_{j}^{0})\right]_{\{a_{k}^{1}\}_{k\in\partial i\setminus j}} & =w(x_{i}^{0})\phi_{i}^{0}(x_{i}^{0})\sum_{\{x_{k}^{0}\}_{k\in\partial i\setminus j}}w(x_{i}^{1}|\boldsymbol{x}_{\partial i}^{0},x_{i}^{0})\prod_{k\in\partial i\setminus j}\psi_{ij}^{0}(x_{k}^{0},x_{i}^{0})\left[A_{k\to i}^{0}(x_{k}^{0},x_{i}^{0})\right]_{a_{k}^{1}}\\
		\left[B_{i\to j}^{t}(x_{i}^{t+1},x_{i}^{t},x_{j}^{t})\right]_{\{a_{t}^{k},a_{t+1}^{k}\}_{k\in\partial i\setminus j}} & =\phi_{i}^{t}(x_{i}^{t})\sum_{\{x_{k}^{t}\}_{k\in\partial i\setminus j}}w(x_{i}^{t+1}|\boldsymbol{x}_{\partial i}^{t},x_{i}^{t})\prod_{k\in\partial i\setminus j}\psi_{ik}^{t}(x_{k}^{t},x_{i}^{t})\left[A_{i\to j}^{t}(x_{i}^{t},x_{j}^{t})\right]_{a_{k}^{t},a_{k}^{t+1}}\\
		& \quad\forall t\in\{1,\ldots,T-1\}\\
		\left[B_{i\to j}^{T}(x_{i}^{T},x_{j}^{T})\right]_{\{a_{k}^{T}\}_{k\in\partial i\setminus j}} & =\phi_{i}^{T}(x_{i}^{T})\sum_{\{x_{k}^{T}\}_{k\in\partial i\setminus j}}\prod_{k\in\partial i\setminus j}\psi_{ik}^{T}(x_{i}^{T},x_{k}^{T})\left[A_{k\to i}^{T}(x_{k}^{T},x_{i}^{T})\right]_{a_{k}^{T}}.
	\end{aligned}
\end{equation}

%\bibliography{references}
%
%\end{document}

\end{document}